\documentclass[aps,prc,floatfix,linenumbers]{revtex4}
\usepackage{amsmath,amssymb,amsfonts}
\usepackage{graphicx}
\usepackage{hhline}
\usepackage{bm}
\usepackage{amsmath}
\usepackage{epsfig}
\usepackage{graphicx}
\usepackage{multirow}

\usepackage{tabularx}
\usepackage{lscape}
\usepackage{rotating}
\usepackage{pstricks}

\usepackage{footnote}
\usepackage{hyperref}

\usepackage{pdfpages}
\usepackage{color} 

\usepackage{enumitem}

\newcommand{\bfvec}[1]{\hbox{\boldmath$#1$\unboldmath}}

\begin{document}
\title{Mechanical properties of the nucleon in the chiral confining model\\ I - formal developments }

\author{Guy Chanfray$^{1}$, Hubert Hansen$^{1}$,  Bikram Keshari Pradhan $^{1}$}

\affiliation{$^{1}$ University Claude Bernard Lyon 1, CNRS/IN2P3, IP2I Lyon, UMR 5822, 69622  Villeurbanne, France }

\begin{abstract}
We discuss the issue of the mechanical stability of the nucleon within a class of models in which massive constituent quarks are subject to a confining potential and are coupled to a surrounding pion cloud enveloping the quark core. The nucleon trial states (either localized factorized wave functions or momentum-projected states) are determined by imposing the von Laue stability condition. This article is primarily devoted to the formal aspects related to the detailed expressions for the total energy (mass), the average pressure, the energy density and the pressure distribution inside the nucleon. It will be accompanied by a complementary article addressing the evolution of nucleon properties with density, associated with the restoration of chiral symmetry.
\end{abstract}

\maketitle
\section{Introduction: trial nucleon states }\label{Intro}

The aim of this paper is to study the stability of the nucleon in a class of chiral models in which the nucleon consists of a quark core generated by a string-like confining interaction, surrounded by a pion cloud coupled to the quarks via the standard pseudo-vector coupling. The effective quark mass is supposed to be generated by an underlying chiral symmetry-breaking mechanism, such as the one based on the Nambu-Jona-Lasinio (NJL) model. This first paper is devoted to the formal development, and a forthcoming accompanying paper \cite{companion}, hereafter referred to as II, will be devoted to the specific case of the chiral confining model (CCM) as described in \cite{Universe,Chanfray2024,Universe2025} to address the question of the density evolution of the nucleon properties.\\

These models are described by the following generic Lagrangian: 
\begin{eqnarray}
 \mathcal{L} &=& \mathcal{L}_{\pi Q}\,+\,\mathcal{L}_C\nonumber\\
 &=& \bar{q}(x)\left[i\,\gamma^{\mu}\partial_\mu-M_q+\frac{1}{2F_\pi}\gamma^j\gamma^5\partial_j\vec{\Phi}\cdot\vec{\tau}\right](x)q(x)
 \,-\,\frac{1}{2}M_\pi^2\,\vec{\Phi}^2 \,+\, \frac{1}{2}  \,\partial^{\mu}\vec{\Phi}\cdot\partial_{\mu}\vec{\Phi}\nonumber\\
 &&\,-\,q^\dagger(x)\,
  W_C\left({\bf r})\,\right)q(x) ,\label{LAGEFF}
\end{eqnarray}
where $q(x)\equiv q({\bf{r}},t)$ is the quark field,  $\vec{\Phi}(x)\equiv\vec{\Phi}({\bf{r}},t)$ is the pion field whose conjugate momentum is denoted by $\vec{\Pi}(x)$. $M_q$ is the constituent quark mass, $F_\pi$ the pion decay constant and $M_\pi$ is the pion mass. Using standard notations, these field operators defined in a large box of volume $V$ can be expanded in terms of  annihilation and creation operators satisfying (discrete) canonical  (anti)commutation relations,
\begin{eqnarray}
q({\bf{r}},t)&=&\frac{1}{\sqrt{V}}\,	\sum_{k=(\bfvec{k},s)}\,\eta_k\,C_k(t)\,e^{i\,\bf{k}\cdot\bf{r}}~, \qquad \eta_k= u({\bf k}, s)=\sqrt{\frac{1}{2}}\left(
\begin{array}{c}
                \sqrt{1+s_k}\,\,\chi_s \\
                \sqrt{1-s_k}\,\,\bfvec{\sigma}\cdot {\bf k}\,\,\chi_s
\end{array}\right)~,\label{QEXP}  \\ 
\Phi_a(\bfvec{r},t)&=&\sum_{\bfvec{k}}\, \frac{1}{\sqrt{2\,\omega_k V}} \,e^{i\bfvec{k}\cdot\bfvec{r}}  \,\left(b_{\bfvec{k} a}(t)\,+\,b^\dagger_{-\bfvec{k} a}(t)\right)~,\label{PIONFIELD}\\\
\Pi_a(\bfvec{r},t)&=&\sum_{\bfvec{k}}\, \frac{-i \omega_k}{2\,\omega_k V} \,e^{i\bfvec{k}\cdot\bfvec{r}}  \,\left(b_{\bfvec{k} a}(t)\,-\,b^\dagger_{-\bfvec{k} a}(t)\right),\label{PIONCONJ}
\end{eqnarray}
with $s_k=M_q/E_k=M_q/\sqrt{k^2+M_q^2}$. Notice that, in the expansion of the quark field, the operators $C$ may formally represent either a quark annihilation operator $B$ or an antiquark creation operator $D^\dagger$. These antiquark states are, however, omitted since they are not required for the present study. In this paper, we will keep the value of these parameters to their vacuum values most of the time, as in the NJLSet1 set of parameters used in \cite{Universe2025}, namely $M_q=356.7\, \rm{MeV}$,\  $M_\pi=140\,\ \rm{MeV}$, $F_\pi=91.7\,\ \rm{MeV}$. The confining potential $W_C(r)$ might be quadratic, linear, or even deformed into a square well potential to mimic the situation of the MIT bag model or the cloudy bag model (CBM). Here we will use, most of the time, the confining potential whose explicit form was proposed in \cite{Universe,Universe2025} and  inspired by the field correlator method (FCM) \cite{Tjon2000,Simonov2002,KNR2017,Simonov2019},
\begin{eqnarray}
 W_C\left({\bf r}\right)&=&  \frac{\sigma}{2\,\sqrt{\pi}\,T_g}\,{\bf r}^2
\,\int_0^1 dv\,\int_0^1 dw\,e^{-\left(\left(v-w\right)^2\frac{{\bf r}^2}{4 T^2_g}\right)}\,-\,\frac{2\,\sigma\,T_g}{\sqrt{\pi}},\label{VCONF}\nonumber\\
&=&\sigma\left(r -\frac{2\,T_g}{\sqrt{\pi}} +\frac{2\,T_g}{\sqrt{\pi}}  e^{-\left(\frac{r^2}{2\,T^2_g}\right)}\,-\, r\,{\rm erfc}\left(\frac{r}{2 T_g}\right)\right)\,-\,\frac{2\,\sigma\,T_g}{\sqrt{\pi}},\label{FCMPOT} 
\end{eqnarray}
where $\sigma$ is the string tension and $T_g$ is the gluon correlation length, itself related to the gluon condensate $\mathcal{G}_2$ according to $T_g=\sqrt{9\sigma/\pi^3 \mathcal{G}_2}$. In the context of the FCM approach, the distance \(r\) represents the distance to the string junction point, which is usually denoted as $\bfvec{X}_J=\bfvec{r}_0$. This confining interaction (without the constant shift) is displayed in Figures 6 and 7 of Ref. \cite{Universe} for typical values of the QCD parameters. It has a quadratic behavior at short distance ($r\ll T_g \sim 0.3\,fm$), whereas, at large distance, it has a linear behavior: $W_c \sim \sigma r$. There is an overall downward shift in the confining potential that creates a pocket, which is particularly welcome as it helps avoid a too-large nucleon mass. The depth of the pocket is of the order $ V_0\sim \sigma T_g\sim 250\,\rm{MeV}$, similar to the constituent quark mass, $M_q \geq \sigma T_g$, generated by the chiral symmetry breaking. This is of course not accidental, since it is the big virtue of this approach that incorporates both chiral symmetry breaking and confinement at the same time. Intending to address later the virial theorem and the von Laue condition for the mechanical stability of the nucleon, we introduce a quantity that will be useful to compute the pressure. We call this quantity a ``confining pressure potential,” which turns out to have a very simple expression:
\begin{equation}
 \bfvec{r}\cdot \bfvec{\nabla}_{\bfvec{r}}W_C(\bfvec{r})=r\,W'_C(r)\,= \,\sigma\,r\,{\rm erf}\left(\frac{r}{2 T_g}\right). \label{eqCPP}
\end{equation}
\\
 The color singlet nucleon state (ignoring for simplicity its explicit spin-color-flavor structure) is generically represented as :
\begin{equation}  
\left|N(\Phi)\right\rangle = V^{3/2}\,\int \frac{d^3 p_1}{(2\pi)^3}\frac{d^3 p_2}{(2\pi)^3}\frac{d^3 p_3}{(2\pi)^3}
\,\Phi\left({\bf p}_1, {\bf p}_2, {\bf p}_3\right)\,B^\dagger_{{\bf p}_1} B^\dagger_{{\bf p}_2} B^\dagger_{{\bf p}_3} \left|0\right\rangle, \label{NUCWF}
\end{equation}
\begin{equation}
\int \frac{d^3 p_1}{(2\pi)^3}\frac{d^3 p_2}{(2\pi)^3}\frac{d^3 p_3}{(2\pi)^3}\,|\left|\Phi\left({\bf p}_1, {\bf p}_2, {\bf p}_3\right)\right|^2 =1,\label{NORM1} 
\end{equation}
where  $\left|0\right\rangle$ is the vacuum of the creation operators $B^\dagger_{{\bf p}_1}$, (i.e., $B_{{\bf p}_1}\left|0\right\rangle=0$) of constituent quark of mass $M$.  From the normalization condition on the Fock-space amplitude, $\Phi\left({\bf p}_1, {\bf p}_2, {\bf p}_3\right)$, this state is automatically  normalized to 1 ($\left\langle N|N\right\rangle =1$) in a large box of volume $V$.\\
In the following sections, we consider different trial nucleon states and, in each case, subsequently describe the local quark densities therein.

\subsubsection{Localized nucleonic state}\label{LocNState}
 We will first consider trial nucleon states where the three-quark Fock-space amplitude is just the product of the three single-quark amplitudes properly projected onto the color singlet state with $I=J=1/2$, namely,
\begin{eqnarray}
\Phi_{\bfvec{X}_N}\left({\bf p}_1, {\bf p}_2, {\bf p}_3\right)&=&e^{-i\,(\bfvec{p}_1 +\bfvec{p}_2+\bfvec{p}_3)\cdot\bfvec{X}_N}\,\, \Phi\left({\bf p}_1\right)\,  \Phi\left({\bf p}_2\right)\,  \Phi\left({\bf p}_3\right)\\
\Phi\left({\bf p}_1\right)\,  
\Phi\left({\bf p}_2\right)\,  
\Phi\left({\bf p}_3\right)&=&R_0(p_1) R_0(p_2) R_0(p_3)\label{WFNOCM},
\end{eqnarray}
with the normalization condition: 
\begin{equation}
\int \frac{d^3 p}{(2\pi)^3}\,\left|\Phi\left({\bf p}\right)\right|^2 =1.\label{NORM2}   
\end{equation}
Such a state, denoted by  $|N(\bfvec{X}_N)\rangle$, represents a localized object with fixed center-of-mass (CM) position $\bfvec{X}_N$, which, in the context of the CCM model discussed in \cite{Universe2025}, can be identified with a string junction position. In this case, the nucleon wave function is just the product of three single-quark orbitals according to:
\begin{eqnarray}
 \left(\Psi_{\bfvec{X}_N}\right)\left(\bfvec{r}_1,\bfvec{r}_2,\bfvec{r}_3\right)  &=&\left\langle 0\right| q(\bfvec{r}_1)q(\bfvec{r}_2)q(\bfvec{r}_3)\left|N(\bfvec{X}_N)\right\rangle=\left(\psi\right)(\bfvec{r}_1 - \bfvec{X}_N)\left(\psi\right)(\bfvec{r}_2 -\bfvec{X}_N)\left(\psi\right)(\bfvec{r}_3 -\bfvec{X}_N),\label{LOCWF}
 \end{eqnarray}
 where the single-quark wave function has a specific quasi-plane-wave (QPW) form:
 \begin{eqnarray}
 \left(\psi\right)({\bf{r}})&=&\int\frac{d^3 p}{(2\pi)^3} \,  e^{-i {\bf p}\cdot {\bf r}}\, (\tilde\psi)({\bf{p}})=\left(
\begin{array}{c}
	u(r)\,\chi_s\\
	i\,\bfvec{\sigma}\cdot \hat{\bf r}\,\,v(r)\,\chi_s
\end{array}\right)\label{QPW2}\\
 (\tilde\psi)({\bf{p}})   &=&\sqrt{\frac{1}{2}}\left(
\begin{array}{c}
	\sqrt{1+s_p}\,\,\hat{\Phi}({\bf{p}})\\
	\sqrt{1-s_p}\,\,\bfvec{\sigma}\cdot \hat{\bf p}\,\,\hat{\Phi}({\bf{p}})
\end{array}\right)
=\sqrt{\frac{E_p + M_q}{2 E_p}}\left(
\begin{array}{c}
	\hat{\Phi}({\bf{p}})\\
	\frac{\bfvec{\sigma}\cdot {\bf p}}{E_p + M_q}\,\,\hat{\Phi}({\bf{p}})
\end{array}\right)\label{QPW1}\\
u(r)&=&\int\frac{d^3 p}{(2\pi)^3} \,j_0(pr) u(p)= \int\frac{d^3 p}{(2\pi)^3} \,j_0(pr) \,\sqrt{\frac{1}{2}}\,   \sqrt{1+s_p}\,\,\Phi(p)\label{QPW3}\\
v(r)&=&\int\frac{d^3 p}{(2\pi)^3} \,j_1(pr) v(p)= \int\frac{d^3 p}{(2\pi)^3} \,j_1(pr) \,\sqrt{\frac{1}{2}}\,   \sqrt{1-s_p}\,\,\Phi(p).\label{QPW4}
\end{eqnarray}
In the following, we will use the following notations:
\begin{equation}
|N:\bfvec{0}_X\rangle\,\equiv\,|N(\bfvec{X}_N=0)\rangle.\label{NOTATIONX}
\end{equation}
In practice, we will use a  Gaussian trial single-quark wave function:
\begin{equation}
\hat{\Phi}({\bf{p}})= \Phi(\bfvec{p})\,\chi_s, \qquad\Phi(\bfvec{p})= R_0(p)= (2\pi)^{\frac{3}{2}}\,\left(\frac{b^2}{\pi}\right)^{\frac{3}{4}}\,e^{-b^2\,p^2/2}.\label{TRIALWF0}  
\end{equation}
\subsubsection{Momentum projected nucleon state}\label{MomentumState}
A significant source of improvement in the description of nucleon properties is to replace the localized nucleon state, not possessing a well-defined total 3-momentum, by a momentum-projected nucleon state with well defined momentum $\bfvec{P}$, denoted by $|N(\bfvec{P}_N)\rangle$, or sometimes for short by $|\bfvec{P}_N\rangle$ in the formal developments. As in Ref. \cite{Gastao}, this amounts to enriching the factorized Fock-space amplitude by the appropriate projection according to: 
\begin{equation}
\Phi_{\bfvec{P}_N}\left({\bf p}_1, {\bf p}_2, {\bf p}_3\right)=\frac{1}{V^{1/2}} N_{\bf{P}_N} \Phi\left({\bf p}_1\right)\,
\Phi\left({\bf p}_2\right)\,  
\Phi\left({\bf p}_3\right)\,(2\pi)^3\,\delta \left(\bf{P}_N - \bf{p}_1 -\bf{p}_2 - \bf{p}_3 \right)\label{WFCM}.
\end{equation}  
 $N_{\bf{P}_N}$ is a normalization factor, whose explicit expression is 
$$N_{\bf{P}_N}=\left(\frac{3}{4\pi b^2}\right)^{3/4}\,exp\left(\frac{P^2_N\, b^2}{6}\right),$$ 
when using the Gaussian trial ansatz of Eq.~\eqref{TRIALWF0}. In the notation $\bfvec{P}_N$, the subscript $N$ indicates that the state is normalized and carries momentum $\bfvec{P}$. Hence, it is implicitly understood that $\bfvec{P}_N\equiv\bfvec{P}$.\\
In this case, the nucleon wave function reads:
\begin{eqnarray}
 \left(\Psi_{\bfvec{P}_N}\right)\left(\bfvec{r}_1,\bfvec{r}_2,\bfvec{r}_3\right)  &=&\left\langle 0\right| q(\bfvec{r}_1)q(\bfvec{r}_2)q(\bfvec{r}_3)\left|N(\bfvec{P}_N)\right\rangle
 \nonumber\\
 &=&\frac{N_{\bf{P}_N}}{\sqrt{V}} \int d^3X\,e^{i\bfvec{P}\cdot\bfvec{X}}\,\left(\psi\right)(\bfvec{r}_1 - \bfvec{X})\left(\psi\right)(\bfvec{r}_2 -\bfvec{X})\left(\psi\right)(\bfvec{r}_3 -\bfvec{X})\nonumber\\
 &\equiv&\frac{1}{\sqrt{V}}e^{i\bfvec{P}\cdot\bfvec{R}}\,\left(\tilde\Psi_{int}\right)(\bfvec{r}_1 - \bfvec{R},\bfvec{r}_2 - \bfvec{R},\bfvec{r}_3 - \bfvec{R}\,;\,\bfvec{P}_N)\nonumber\\
 \left(\tilde\Psi_{int}\right)&=& N_{\bf{P}_N}\,\int d^3Y\,e^{i\bfvec{P}\cdot\bfvec{Y}}\,\left(\psi\right)(\bfvec{x}_1 - \bfvec{Y})\left(\psi\right)(\bfvec{x}_2 -\bfvec{Y})\left(\psi\right)(\bfvec{x}_3 -\bfvec{Y}).
 \end{eqnarray}
Here $\bfvec{R}=(\bfvec{r}_1 +\bfvec{r}_2 +\bfvec{r}_3)/3$ is the position of the center of mass of the three-quark system and $\bfvec{x}_1 =\bfvec{r}_1 -\bfvec{R},\,\bfvec{x}_2 =\bfvec{r}_2 -\bfvec{R},\,\bfvec{x}_3 =\bfvec{r}_3 -\bfvec{R}$ represent the position of each of the quarks relative to their center of mass. These relative positions only depend on the Jacobi coordinates  $\bfvec{\rho}=\bfvec{r}_1 -\bfvec{r}_2$ and $\bfvec{\lambda}=\bfvec{r}_3 - (\bfvec{r}_1 +\bfvec{r}_2)/2$. One sees that this wave function is completely delocalized and appears as the product of a plane wave associated with the total momentum $\bfvec{P}$ carried by the center of mass and an intrinsic wave function in which the internal structure of the nucleon is encoded. This wave function is translationally invariant in the sense that it consists of a CM plane wave multiplied by an intrinsic wave function that does not depend on the CM position $\bfvec{R}$. In the following, we will use the following notation:
\begin{equation}
   |N:\bfvec{0}_P\rangle\,\equiv\,|N(\bfvec{P}_N=0)\rangle .\label{NOTATION}
\end{equation}
Also note that the discrete $\left|N(\bfvec{P}_N)\right\rangle$ states satisfying $\left\langle \bfvec{P}'_N |\bfvec{P}_N\right\rangle=\delta_{\bfvec{P}'_N,\bfvec{P}_N}$ are related to the usual momentum eigenstates of non-relativistic quantum mechanics which obey $\left\langle \bfvec{P}' |\bfvec{P}\right\rangle=(2\pi)^3\delta^{(3)}(\bfvec{P}'-\bfvec{P})$, through the relation $\left|\bfvec{P}\right\rangle=\sqrt{V} \left|\bfvec{P}_N\right\rangle$.
\subsubsection{Local quark densities}\label{LOCAL}
Let us consider the local densities, $t_\Gamma(\bfvec{r})\equiv t_\Gamma(x=(0,\bfvec{r}))$ (i.e., implicitly taken at $t=0$ in this discussion),  associated with local density operators, $\hat{T}_\Gamma(\bfvec{r})=\bar{q}(\bfvec{r})\,\Gamma \,q(\bfvec{r})$. Examples include the quark kinetic energy density operator, $\mathcal{H}_{QK}(\bfvec{r})=\bar{q}(\bfvec{r})\,(-i\bfvec{\gamma}\cdot\bfvec{\nabla}_{\bfvec{r}}) \,q(\bfvec{r})$, the scalar density operator, $\hat{\rho}_S(\bfvec{r})=\bar{q}(\bfvec{r})\,q(\bfvec{r})$,  or the quark number density operator, $\hat{\rho}_B(\bfvec{r})=\bar{q}(\bfvec{r})\,\gamma_0 \,q(\bfvec{r})$.\\
In the case of a localized nucleon state (see \ref{LocNState}), a very natural definition of these densities is obtained by just taking the matrix element of the local density operator, i.e., 
\begin{equation}
t_\Gamma(\bfvec{r})=\left\langle 0_X\right|\,\hat{T}_\Gamma(\bfvec{r})\,  \left|0_X\right\rangle =3\,(\bar{\psi})(\bfvec{r})\,\Gamma\,(\psi)(\bfvec{r}).
\end{equation}\\
In the case of a pure momentum eigenstate (see \ref{MomentumState}), this definition makes no physical sense, as translational invariance leads to a uniform density. Hence, one has to find a way to localize the quantum system. Following  Wigner \cite{Wigner}, as clearly summarized in \cite{Lorce2019}, a sensitive way consists in  replacing the pure momentum eigenstate $\left|\bfvec{P}_N\right\rangle$ with a localized nucleonic system with average momentum $\bfvec{P}_N=\bfvec{P}$ and average position $\bfvec{R}_N=\bfvec{R}$ described by the phase space density operator:
\begin{equation}
 \hat{\rho}_{\bfvec{R},\bfvec{P}} =\int \frac{V d^3\Delta}{(2\pi)^3}\, e^{-i\bfvec{\Delta}\cdot\bfvec{R}}\,\left|\bfvec{P}_N -\frac{\bfvec{\Delta}_N}{2}\right\rangle\left\langle \bfvec{P}_N +\frac{\bfvec{\Delta}_N}{2}\right|.            
\end{equation}
The local density $t_\Gamma(\bfvec{r})$ is thus assumed to be given by the expectation value of the operator $\hat{T}_\Gamma (\bfvec{r})$  at the spatial point $\bfvec{r}$ in the statistical ensemble described by the above density operator:
\begin{eqnarray}
t_\Gamma (\bfvec{r}; \bfvec{P};  \bfvec{R})&=&Tr\left[\hat{\rho}_{\bfvec{R},\bfvec{P}}\hat{T}_\Gamma (\bfvec{r})\right] =   \int \frac{V d^3\Delta}{(2\pi)^3}e^{-i\bfvec{\Delta}\cdot\bfvec{R}}\,\left\langle \bfvec{P}_N +\frac{\bfvec{\Delta}_N}{2}\right|\hat{T}_\Gamma (\bfvec{r}) \left|\bfvec{P}_N -\frac{\bfvec{\Delta}_N}{2}\right\rangle\nonumber\\
&=&\int \frac{d^3\Delta}{(2\pi)^3}e^{-i\bfvec{\Delta}\cdot\bfvec{R}}\,\left\langle \bfvec{P} +\frac{\bfvec{\Delta}}{2}\right|\hat{T}_\Gamma (\bfvec{r}) \left|\bfvec{P} -\frac{\bfvec{\Delta}}{2}\right\rangle\equiv \left[\hat{T}_\Gamma (\bfvec{r})\right]_{WT}(\bfvec{R},\bfvec{P}).\label{DEFWIG}
\end{eqnarray}
Hence, the local density appears as the Wigner transform of the local density operator taken at the phase-space point $(\bfvec{R},\bfvec{P})$ corresponding to the average position and momentum of the nucleonic system. As a side remark, this notion of the Wigner transform is widely used in the semi-classical approaches in nuclear physics \cite{ChanfrayPRA,Schuck89}.
An important consistency check is that, thanks to translational invariance,  the volume integral of the density thus defined correctly reproduces the value of the associated ``charge" (i.e., the spatially integrated quantity), which is itself well-defined for the pure momentum state,
\begin{equation}
 \int d^3r\,  t_\Gamma (\bfvec{r}; \bfvec{P};  \bfvec{R})= V\,\left\langle \bfvec{P}_N\right|\hat{T}_\Gamma (0) \left| \bfvec{P}_N \right\rangle=\int d^3r\, \left\langle \bfvec{P}_N\right|\hat{T}_\Gamma (\bfvec{r}) \left| \bfvec{P}_N\right\rangle,
\end{equation}
which does not depend on $\bfvec{R}$ as expected. Since the physics should not depend on $\bfvec{R}$, one could always localize the nucleonic system around $\bfvec{R}=0$. For a nucleon at rest, i.e., $\bfvec{P}=0$, the matrix element of the local density operator in the Wigner sense is given by: 
\begin{eqnarray}
\left\langle 0_P  \right| \hat{T}_\Gamma (\bfvec{r}) \left|0_P\right\rangle_W &\equiv&  t_\Gamma (\bfvec{r})=Tr\left[\hat{\rho}_{\left(\bfvec{R}=0,\bfvec{P}=0\right)}\,\hat{T}_\Gamma (\bfvec{r})\right] \nonumber\\
&=&\int \frac{V d^3\Delta}{(2\pi)^3}\, \left\langle \bfvec{\Delta}_N/2\right|\hat{T}_\Gamma (\bfvec{r}) \left| -\bfvec{\Delta}_N/2\right\rangle=\int \frac{V d^3\Delta}{(2\pi)^3}\,e^{-i\bfvec{\Delta}\cdot\bfvec{r}} \left\langle \bfvec{\Delta}_N/2\right|\hat{T}_\Gamma (0) \left| -\bfvec{\Delta}_N/2\right\rangle\label{WIGDEF}
\end{eqnarray}
The matrix element of the local density operator, $\left\langle \bfvec{\Delta}_N/2\right|\hat{T}_\Gamma (0) \left| -\bfvec{\Delta}_N/2\right\rangle$, evaluated between momentum states differing by $\bfvec{\Delta}$, may be promoted to the status of a physical observable. Indeed, it can be interpreted as a Sachs form factor \cite{RGS} expressed in the Breit frame (in analogy to the electromagnetic form factors
\cite{Xiong}) that may be experimentally accessed through scattering processes involving an external probe coupled to the nucleon via the local density operator $\hat{T}_\Gamma (\bfvec{r})$. This is indeed the case for the components of the energy momentum tensor (EMT), whose matrix elements, $\left\langle \bfvec{\Delta}_N/2\right|\hat{T}^{\mu\nu}(0) \left| -\bfvec{\Delta}_N/2\right\rangle$,  can be obtained, via a modeling of the gravitational form factors,  from the generalized parton distributions (GPDs) \cite{Polyakov,Lorce2019}, which are accessible in several exclusive processes, such as deeply virtual Compton scattering \cite{DVCS} and meson production \cite{MESPROD}. The components of the associated densities that constitute the static EMT tensor can thus be extracted from the data.
At this point, a comment may be in order if we recall that a spatial interpretation and the definition of local densities require a prescription, since an exact momentum eigenstate is completely delocalized. The “semiclassical” prescription, defined in terms of a Wigner transform  (Eq. \eqref{DEFWIG}) of the corresponding local operator, has the advantage of leading to a stress tensor that can be related to observable form factors. In our opinion, this observability certainly reinforces the physical status of  $\varepsilon (r)=t^{00}(r)$ interpreted as a genuine energy density and $p(r)=t^{ii}(r)/3$ interpreted as a genuine mechanical pressure distribution inside the nucleon.\\
To conclude this discussion relative to the physical nucleon states with well defined 3-momentum, let us note that all the contributions to the Hamiltonian matrix elements (and, more generally, to the energy–momentum tensor) that originate from a translationally invariant Lagrangian do not require the introduction of local (Wigner) densities. However—and this is a crucial point—the confining part of the interaction, $H_C=\int d^3 r\,\mathcal{H}_{C}(\bfvec{r})=\int d^3 r\,W_C(r)q^\dagger(\bfvec{r})q(\bfvec{r})$, which breaks translational invariance, does require a proper definition of the quark number distribution. In reality, the confining interaction experienced by a quark depends on its distance from the center of force, which, in the CCM model, coincides with the position of the string junction, conventionally taken to be at the origin. We therefore assume that the position of the string junction coincides with the mean position of the quasi-localized nucleonic system. Hence, the  matrix element in the Wigner sense of the confining interaction is:
\begin{eqnarray}
\left\langle 0_P  \right| H_C \left|0_P\right\rangle_W &=&\int d^3r\, \left\langle 0_P  \right| \mathcal{H}_C(\bfvec{r}) \left|0_P\right\rangle_W =\int d^3r\,W_C(r) \left\langle 0_P  \right| \hat{\rho}_B(\bfvec{r}) \left|0_P\right\rangle_W \nonumber\\
&=&
\int d^3r\,W_C(r)\,Tr\left[\hat{\rho}_{\left(\bfvec{R}=0,\bfvec{P}=0\right)}\,\hat{\rho}_B(\bfvec{r})\right]
= \int d^3r\,W_C(r)\int \frac{V d^3\Delta}{(2\pi)^3}\, \left\langle \bfvec{\Delta}_N/2\right|q^\dagger(\bfvec{r})q(\bfvec{r}) \left| -\bfvec{\Delta}_N/2\right\rangle\nonumber\\
&=& 3\int d^3r\,W_C(r)\rho_B(r).  
\end{eqnarray}
When using the Gaussian trial single-quark Fock-space amplitude, one can demonstrate analytically that the quark number density  in the Wigner sense, $\rho_B(r)$, normalized to one, is given by:
\begin{eqnarray}
3\,\rho_B(r)&=&\int\frac{V d^3\Delta}{(2\pi)^3}\, \langle \bfvec{\Delta}_N/2|\,q^\dagger(\bfvec{r})\,q(\bfvec{r})|-\bfvec{\Delta}_N/2\rangle\nonumber\\
&=&3\,\int\frac{d^3\Delta}{(2\pi)^3}\, e^{-i\,\bfvec{\Delta} \cdot\bfvec{r}}\, e^{5 b^2\Delta^2/24}\,\int d^3x\,e^{i\,\bfvec{\Delta} \cdot\bfvec{x}} \,\left(\psi\right)^\dagger_E({\bf{x}})\,\left(\psi\right)_E({\bf{x}})\\
&=&3\,\int\frac{d^3\Delta\,d^3x}{(2\pi)^3}\, e^{5 b^2\Delta^2/24}\,j_0(\Delta r)\,j_0(\Delta x)\,
\left(\psi^\dagger_E \psi_E\right)(x),
\end{eqnarray}
with the normalization condition:
\begin{equation}
\int d^3 r  \rho_B(r) =1.
\end{equation}
Here $\left(\bar{\psi}\right)_E({\bf{x}})$ is the QPW wave function defined in \eqref{QPW2},\eqref{QPW1},\eqref{QPW3},\eqref{QPW4},\eqref{TRIALWF0}, but with the  parameter $b$ replaced with an effective parameter $b_E$ such that:
\begin{equation}
 \left(\psi\right)({\bf{x}}) \,\to\,   \left(\psi\right)_E({\bf{x}})\qquad\Leftrightarrow\qquad b^2\,\to\, b^2_E=\frac{3}{2}\,b^2.
\end{equation}
 The root-mean-square radius, $\sqrt{\langle r^2_B\rangle}$, of the quark core baryon number distribution is defined through the baryonic form factor:
 \begin{equation}
  \langle r^2_B\rangle=-6\left[\frac{F_B(q^2)}{dq^2}\right]_{q^2=0}\qquad\hbox{with}\qquad F_B(q^2)= \int d^3r\,e^{i\,\bfvec{q}\cdot\bfvec{r}}  \rho_B(r).
 \end{equation}
 For the case of the localized nucleon state, it simply reads:
 \begin{equation}
|N:\bfvec{0}_X\rangle:\qquad\qquad     \langle r^2_B\rangle=\int dr\,4\pi r^2\,r^2\,\left(u^2(r)+v^2(r)\right),
 \end{equation}
 whereas, for the case of a momentum projected nucleon state with a Gaussian Fock-space amplitude, it is modified according to:
 \begin{equation}
 |N:\bfvec{0}_P\rangle:\qquad\qquad  \left\langle r^2_B\right\rangle =  \int dr\,4\pi r^2\,\left(u_E^2(r)+v_E^2(r)\right)\,
 \,-\,\frac{5}{4}\,b^2=\left\langle r^2_B\right\rangle_E \,-\,\frac{5}{6}\,b_E^2 .\label{RBEX1}
\end{equation}
 The resulting  quark core size can then be related   to the experimentally measured isoscalar radius of the nucleon, $\sqrt{\langle r^2_S\rangle}\simeq 0.77\, \rm{fm}$ \cite{Xiong}, assuming vector dominance phenomenology  since $\sqrt{\langle r^2_S\rangle}$ is connected to the three-pion spectrum dominated by the very narrow $\omega$ meson:
\begin{equation}
  \sqrt{\langle r^2_B\rangle}=\sqrt{\langle r^2_S\rangle\,-\,\frac{6}{m^2_\omega}} =0.46\,\rm{fm}. 
\end{equation}
Let us conclude this first section with some comments on the treatment of translational invariance. 
Indeed, one well-known difficulty of this type of model, with a localized center of confining force that here physically corresponds to the string junction position, is the breaking of translational invariance. A possible way to fully restore translational invariance is to consider the string junction point as a collective dynamical variable  whose position should be fixed by minimization of the interaction between
the three quarks, yielding for the string junction position, $\bf{X}_J=\bf{r}_0$, the so-called Torricelli point. The  resulting interaction is neither an additive one-body nor an additive two-body interaction, but a single-string $Y$ junction, which is of a genuine three-body nature (see Ref. \cite{Simonov2002}). Another often-used possibility, as in Ref. \cite{Gastao},  is to identify the string junction point with the CM position ${X}_J\equiv\bf{R}=(\bf{r}_1 +\bf{r}_2 +\bf{r}_3)/3$, implying that the string junction coordinate is translated and projected together with the quarks. However, these procedures are not well suited to our purpose, especially if one intends to define  a local EMT and the corresponding energy density and pressure distribution. This is one of the main reasons which led us to calculate the matrix element of the confining interaction for a zero momentum state by replacing the projected state represented by the wave packet,
$\left|N(\bf{P}_N=0)\right\rangle=(N_{\bfvec{P_N}}/\sqrt{V})\,\int d^3\,X_N \,e^{i\bf{0}\cdot\bf{X}_N}  \left|N(\bf{X}_N)\right\rangle$,    
with  a phase-space density operator, $\rho_{(\bf{R}=0,\bf{P}=0)}$, given by Eq. \eqref{DEFWIG} representing a localized system with average momentum $\bf{P}=0$ and average position of the string junction point at $\bf{R}=0$, as explained above. This  specific Wigner-based model prescription  for restoring translational invariance is, of course, not equivalent to a fully translationally invariant relativistic three-quark treatment. Actually, it leads, in general,  to a result for the matrix element of the confining interaction that differs  from the one obtained  by directly calculating $\left\langle\sum_j\,W_C(\bf{r}_j\,-\,(\bf{r}_1 +\bf{r}_2 +\bf{r}_3)/3)\right\rangle$ on the projected state, although the two results are expected to be close. Indeed, we have explicitly checked that, for  a diquark system, the two approaches formally give exactly the same result in the non-relativistic limit.
\section{The Hamiltonian: quark coupling to finite size pions}
The Hamiltonian associated with the Lagrangian \ref{LAGEFF} reads:
\begin{eqnarray}
H &=& H_Q\,+\,H_{\pi Q}\,+\,H_{\pi\pi} \label{HTOT}\\   
&\equiv&\int d^3 r\,\left(\mathcal{H}_Q\,+\,\mathcal{H}_{\pi Q}\,+\,\mathcal{H}_{\pi\pi} \right)(\bfvec{r})\label{HDENS0}\\
&\equiv&\int d^3 r\,\left(\mathcal{H}_{QK}\,+\,\mathcal{H}_{QM}\,+\,\mathcal{H}_{C}\,+\,\mathcal{H}_{\pi Q}\,+\,\mathcal{H}_{\pi c} 
\,+\,\mathcal{H}_{\pi K}\,+\,\mathcal{H}_{\pi M}\right)(\bfvec{r})\label{HDENS}\\
H_Q &=& \int d^3 r\, q^\dagger(\bfvec{r})\left[-i\bfvec{\alpha}\cdot\bfvec{\nabla}_{\bfvec{r}}\,+\, \beta M_q \, +  \,\,W_C(\bfvec{r})\right]\,q(\bfvec{r})\equiv H_{QK}\,+\, H_{QM}\,+H_C\label{HQ}\\
H_{\pi Q} &=&- \int d^3 r\,\frac{1}{2 F_\pi}\,\partial_j\vec{\Phi}(\bfvec{r})\cdot \bar{q}(\bfvec{r})\gamma^j\,\gamma_5\,\vec{\tau}q(\bfvec{r})\equiv 
-\int d^3 r\,\frac{1}{2 F_\pi}\,\bfvec{\nabla}\vec{\Phi}(\bfvec{r})\cdot \vec{\bfvec{S}}(\bfvec{r})\label{HPIQ}\\
H_{\pi\pi}&=& \int d^3 r\,\frac{1}{2}\left(\vec{\Pi}^2\,+\,\partial_{j}\vec{\Phi}\cdot\partial_{j}\vec{\Phi}\,+\, M_\pi^2\,\vec{\Phi}^2\right)(\bfvec{r})\equiv H_{\pi c}\,+\,H_{\pi K}\,+\,H_{\pi M},\label{HPI}
\end{eqnarray}
where we have introduced the quark bilinear operator $\vec{\bfvec{S}}(\bfvec{r})$ given below along with its longitudinal Fourier transform:
\begin{equation}
\vec{\bfvec{S}}(\bfvec{r})= \bar{q}(\bfvec{r})\bfvec{\gamma}\,\gamma_5\,\vec{\tau}q(\bfvec{r}),\quad  
 \vec{S}^\dagger_{\bf k} =\int d^3r \,e^{-i{\bf k}\cdot{\bf r}} \,\vec{\bfvec{S}}(\bfvec{r})\cdot{\hat{\bf k}},\quad  
 \vec{S}_{\bf k} =\int d^3r \,e^{i{\bf k}\cdot{\bf r}} \,\vec{\bfvec{S}}(\bfvec{r})\cdot{\hat{\bf k}}.
\end{equation} 
As discussed in \cite{Universe2025}, the model, as it stands, suffers from an instability problem due to  a diverging attractive pion-quark interaction energy (i.e., the pionic self-energy) when the quark core size goes to zero. As the size of the quark core decreases, the negative pion pressure may become very large, surpassing the Fermi pressure and potentially causing collapse of the ``quark core bag". The origin of this rather old problem of chiral bag models \cite{Brown,Pirner} can be identified if one realizes that the elementary pion field couples derivatively  to the quarks primarily at the surface of the quark core (i.e., $r_c\sim b\sim 1/\sqrt{\sigma}\sim 0.5\,\rm{fm}$ ), which is not larger than the physical pion size extracted from the pion electromagnetic form factor, $\sqrt{\langle r^2_\pi\rangle} =0.65\,\rm{fm}$. In the NJL model, this rms radius (keeping only the dominant three-quark loop contribution, see Section VI-4 of Ref. \cite{KLE}) is such that: 
$\langle r^2_\pi\rangle_{NJL}=\frac{3}{4\pi^2 F^2_\pi}=0.34\,\rm{fm}^2\simeq(0.6\, \rm{fm})^2.$ 
Hence, the finite pion size cannot be ignored in the calculation of the pionic self-energy. To incorporate this effect, we replace the local coupling of the elementary pion field with a non-local coupling characterized by a  spreading function $\tilde{P}(x)$ (with Fourier transform $P(t)$) as:
\begin{equation}
 \bfvec{\nabla}\vec{\Phi}(\bfvec{r}) \cdot\vec{\bfvec{S}}({\bfvec{r}})
\quad \to\quad \int d^3x \,\bfvec{\nabla}\vec{\Phi}({\bf r}) \cdot\tilde{P}({\bfvec{r}}-\bfvec{x})\vec{\bfvec{S}}(\bfvec{x}). 
\end{equation} 
 This is equivalent to making the replacements:
\begin{eqnarray}
 \vec{\bfvec{S}}({\bfvec{r}})&\to&  \tilde{\bfvec{S}}(\bfvec{r})= \int d^3x \,\tilde{P}(\bfvec{r}-\bfvec{x})\vec{\bfvec{S}}(\bfvec{x})=\int d^3x \,\int \frac{d^3t}{(2\pi)^3}\,e^{-i{\bf t}\cdot({\bf r}-{\bf x})}\,P({\bfvec{t}})\,\vec{\bfvec{S}}(\bfvec{x})
 \end{eqnarray}
 \begin{eqnarray}
 \vec{S}_{\bf k} =\int d^3r \,e^{i{\bf k}\cdot{\bf r}} \,\vec{\bfvec{S}}(\bfvec{r})\cdot{\hat{\bf k}}&\to&
\tilde{S}_{\bf k}= \int d^3r \,e^{i{\bf k}\cdot{\bf r}} \,\tilde{\bfvec{S}}(\bfvec{r})\cdot{\hat{\bf k}}=P (\bfvec{k})\,\vec{S}_{\bf k}.
\end{eqnarray}
Hence the pion-quark Hamiltonian and Hamiltonian density operators introduced in Eqs. \eqref{HDENS}, \eqref{HPIQ} become:
\begin{equation}
 H_{\pi Q} = \int d^3 r\,\mathcal{H}_{\pi Q}=
-\int d^3 r\,\frac{1}{2 F_\pi}\,\bfvec{\nabla}\vec{\Phi}(\bfvec{r})\cdot \tilde{\bfvec{S}}(\bfvec{r}).\label{HPIQMOD}\\   
\end{equation}
This Hamiltonian keeps a local form associated with the position of the pion field, the non-locality being  hidden in the displaced quark source, $\tilde{\bfvec{S}}(\bfvec{r})= \int d^3x \,\tilde{P}(\bfvec{r}-\bfvec{x})\vec{\bfvec{S}}(\bfvec{x})$.

We take for $\tilde{P}(\bfvec{\eta})$ a  Gaussian form:
$$\tilde{P}(\bfvec{\eta})=\left(\frac{1}{\pi\,\rho^2_\pi}\right)^{3/2}\,exp\left(-\frac{\eta^2}{\rho^2_\pi}\right),\qquad P({\bf{t}})=\int d^3\eta\,e^{-i\bfvec{t}\cdot\bfvec{\eta}}\,\tilde{P}(\bfvec{\eta})= exp\left(-\rho^2_\pi\frac{t^2}{4}\right).$$
We expect the size parameter $\rho_\pi$ to be directly related to the electromagnetic size of the pion. In the framework of an underlying NJL model, since this pion radius arises from a constituent quark loop, we also expect that it depends on the constituent quark mass $M_q$ and consequently on the pion decay constant parameter $F_\pi$ \cite{Pirner}. For orientation, we  take:
\begin{equation}
\rho^2_\pi= \frac{c_\pi}{4\pi^2 F^2_\pi}=c_\pi\,(0.34\, fm)^2\label{PIONSIZE}~,
\end{equation}
where $c_\pi$ is an adjustable parameter expected to be close to unity.\\

The equation of motion for the quark field reads:
\begin{equation}
i\frac{\partial q(\bfvec{r},t)}{\partial t} =h_D(\bfvec{r}) \,q(\bfvec{r},t),\label{HF1}   
\end{equation}
where the single particle Dirac Hartree-Fock Hamiltonian is 
\begin{equation}
h_D(\bfvec{r})=  -i\bfvec{\alpha}\cdot\vec\nabla_{\bfvec{r}}\,+\, \beta  M_q \, -  \,  \frac{1}{2 F_\pi}\,\int d^3x\,\tilde{P}(\bfvec{r}-\bfvec{x})\partial_j\vec{\Phi}(\bfvec{x})\,\gamma^j\,\gamma_5\,\vec{\tau}.\label{HF2}
\end{equation}
It follows that, once the equation of motion is satisfied, the Lagrangian reduces to the pure pionic piece:  
\begin{equation}
\mathcal{L} \rightarrow\mathcal{L}_{\pi\pi} =\partial^{\mu}\vec{\Phi}\cdot\partial_{\mu}\vec{\Phi}\,-\,\frac{1}{2}M_\pi^2\,\vec{\Phi}^2 =\frac{1}{2}\left(\vec{\Pi}^2\,-\,\partial_{j}\vec{\Phi}\cdot\partial_{j}\vec{\Phi}\,-\, M_\pi^2\,\vec{\Phi}^2\right).\label{REDLAG}
\end{equation}
This will be used in practice for the expression of the pressure operator in the next section.
\section{Virial theorem and the von Laue nucleon stability condition}

In this section, we will establish the von Laue stability condition. To achieve this, we  first need to properly define a local energy-momentum tensor (EMT) in the presence of a non-local pion-quark coupling. 

\subsection{Local energy-momentum tensor in presence of a non-local pion-quark coupling}
Let us consider the variation of the action $S=\int d^4x \,\mathcal{L}(x)$ with $\mathcal{L}(x)$  defined in Eq.~\eqref{LAGEFF}, under a translation $x^\nu\to x^\nu +\epsilon^\nu$:
\begin{equation}
\delta S^{(D)}=\int d^4x \,\partial^\nu\mathcal{L} (x) \, \epsilon_\nu. \label{EQSD}
\end{equation}
This variation comes, in part, from the variation of the various fields $\delta\phi_r(x)=\phi_r(x+\epsilon)-\phi_r(x)=\partial^\nu\phi_r(x)\,\epsilon_\nu$, and in part from the variation of the confining potential which breaks translational invariance:
\begin{eqnarray}
 \delta S^{(F)}&=&\int d^4x\,\left[\left(\frac{\partial\mathcal{L}}{\partial \phi_r}-\partial_\mu\left(\frac{\partial\mathcal{L}}{\partial_\mu \phi_r}\right)\right)\partial^\nu\phi_r\,\epsilon_\nu\,+\,\partial_\mu\left(\frac{\partial\mathcal{L}}{\partial_\mu \phi_r}\partial^\nu\phi_r\right) \right]  \nonumber\\
 && -\int d^4x\,q^\dagger(x)\,\partial^\nu W(x)\,q(x)\,\epsilon_\nu \label{DACT1},
\end{eqnarray} 
where the term in the first parenthesis vanishes from the equation of motion. Identifying the two expressions (Eqs. \eqref{EQSD},\eqref{DACT1}) for the variation of the action, we arrive at: 
\begin{eqnarray}
\delta S^{(F)}-\delta S^{(D)}=\int\,d^4x\,\left[\partial_\mu\left(\frac{\partial\mathcal{L}}{\partial_\mu \phi_r}\partial^\nu\phi_r\,-\,\mathcal{L}(x)\,{g}^{\mu\nu}\right) \,-q^\dagger(x)\,\partial^\nu W(x)\,q(x)\right]\,\epsilon_\nu=0.    
\end{eqnarray}
It follows that,
\begin{eqnarray}
 \partial_\mu \hat{T}^{\mu\nu}(x) &=& q^\dagger(x)\,\partial^\nu W(x)\,q(x)\quad\hbox{with}\quad \hat{T}^{\mu\nu}(x)= \sum_{fieds\,\varphi_r}\,\frac{\partial\mathcal{L}}{\partial_\mu \varphi_r}\partial^\nu\varphi_r -\mathcal{L}(x)\,{g}^{\mu\nu},\label{VIRIAL0}
\end{eqnarray}
where $\hat{T}^{\mu\nu}(x)$ is the canonical energy-momentum tensor (EMT), which is not conserved due to the lack of translational invariance induced by the presence of the confining interaction.

If we incorporate the finite pion size, i.e., the non-local coupling of the pion to the quarks in the effective action, the pion-quark coupling action  is modified according to: 
\begin{eqnarray}
S_{\pi Q}=\int d^4x \,\frac{1}{2\,F_\pi}\bfvec{\nabla}\vec{\Phi}(x)\cdot\vec{\bfvec{S}}(x)\to \int d^4x\, d^4y \,\frac{1}{2\,F_\pi}\bfvec{\nabla}\vec{\Phi}(x)\,\tilde{\mathcal{P}}(x-y)\cdot\vec{\bfvec{S}}(y),
\end{eqnarray}
with $\tilde{\mathcal{P}}(x-y)=\delta(x^0-y^0)\tilde{P}(\bfvec{x}-\bfvec{y})$.  The variation of this pion-quark action reads:
\begin{eqnarray}
\delta S_{\pi Q}^{(F)}&=&   \frac{1}{2\,F_\pi} \int d^4x\, \epsilon_\nu\,\partial^\nu \bar{q}(x)\bfvec{\gamma}\,\gamma_5\,\vec{\tau}q(x)\cdot \left(\int d^4 y\,\tilde{\mathcal{P}}(x-y)\bfvec{\nabla}\vec{\Phi}(y) \right)\nonumber\\
&& + \frac{1}{2\,F_\pi} \int d^4x\, \epsilon_\nu\,\bar{q}(x)\bfvec{\gamma}\,\gamma_5\,\vec{\tau}\partial^\nu q(x)\cdot \left(\int d^4 y\,\tilde{\mathcal{P}}(x-y)\bfvec{\nabla}\vec{\Phi}(y) \right)\nonumber\\
&&- \frac{1}{2\,F_\pi}\int d^4x\,\epsilon_\nu\,\partial^\nu\,\vec{\Phi}(x)\cdot \left(\bfvec{\nabla}_x\cdot\int d^4 y\,\tilde{\mathcal{P}}(x-y)\vec{\bfvec{S}}(y)\right)\nonumber\\
&&+\frac{1}{2\,F_\pi}\int d^4x\, \epsilon_\nu\,\bfvec{\nabla}_x\cdot\left(\partial^\nu\vec{\Phi}\cdot\int d^4 y\,\tilde{\mathcal{P}}(x-y)\vec{\bfvec{S}}(y)\right).
\end{eqnarray}
The first three lines are absorbed into the equation of motion for the quark field and the pion field, 
\begin{eqnarray}
&&\left(i\gamma^\mu\partial_\mu\,-\,\beta M_q\,-\,W_C(x) \right) q(x) +\frac{1}{2\,F_\pi}\bfvec{\gamma}\,\gamma_5\,\vec{\tau}q(x)\cdot \left(\int d^4 y\,\tilde{\mathcal{P}}(x-y)\bfvec{\nabla}\vec{\Phi}(y) \right)=0 \label{QEOS}, \\
&&\partial^\nu \partial_\nu\vec{\Phi} \,+\,M^2_\pi\vec{\Phi}+\frac{1}{2\,F_\pi}\bfvec{\nabla}\cdot\left(\int d^4 y\,\tilde{\mathcal{P}}(x-y)\vec{\bfvec{S}}(y)\right)=0,
\end{eqnarray}
and $\delta S_{\pi Q}^{(F)}$ reduces to: 
\begin{equation}
\delta S_{\pi Q}^{(F)}=\frac{1}{2\,F_\pi}\int d^4x\, \epsilon_\nu\,\bfvec{\nabla}_x\cdot\left(\partial^\nu\vec{\Phi}\cdot\int d^4 y\,\tilde{\mathcal{P}}(x-y)\vec{\bfvec{S}}(y)\right).     
\end{equation}
The explicit variation of $S_{\pi Q}$ is:
\begin{eqnarray}
\delta S_{\pi Q}^{(D)}&=& \frac{1}{2\,F_\pi}\int d^4x\, d^4y\,\epsilon_\nu  \big[\partial_x^\nu(\bfvec{\nabla}\vec{\Phi}(x))\cdot\tilde{\mathcal{P}}(x-y)\vec{\bfvec{S}}(y)\nonumber\\
&&\qquad\qquad +\,\bfvec{\nabla}\vec{\Phi}(x))\cdot(\tilde{\mathcal{P}}(x-y)\partial_y^\nu\bfvec{S}(y)\big]\nonumber\\
&=&\frac{1}{2\,F_\pi}\int d^4x\ d^4 y\,\epsilon_\nu\partial^\nu \left(\bfvec{\nabla}\vec{\Phi}(x))\cdot\tilde{\mathcal{P}}(x-y)\vec{\bfvec{S}}(y))\right).
\end{eqnarray}
It follows that:
\begin{eqnarray}
\delta S_{\pi Q}^{(F)}-\delta S_{\pi Q}^{(D)} &=&\int d^4x\,\left[\frac{1}{2\,F_\pi}\bfvec{\nabla}\cdot\left(\partial^\nu\vec{\Phi}(x)\cdot\tilde{\bfvec{S}}(x)\right)\,-\,\frac{1}{2\,F_\pi}\partial_\mu\left(g^{\mu\nu}\bfvec{\nabla}\vec{\Phi}(x))\cdot\tilde{\bfvec{S}}(x)\right)\right]\epsilon_\nu\nonumber\\
&=&\int d^4x\,\frac{1}{2\,F_\pi}\partial_\mu\left[\frac{\partial}{\partial_\mu \Phi_b(x)}\left(\bfvec{\nabla}\vec{\Phi}(x))\cdot\tilde{\bfvec{S}}(x)\right)\partial^\nu\Phi_b(x)\, -\,\left(g^{\mu\nu}\bfvec{\nabla}\vec{\Phi}(x))\cdot\tilde{\bfvec{S}}(x)\right)  \right]\epsilon_\nu\nonumber\\
&\equiv&\int d^4x\,\partial_\mu \hat{T}^{\mu\nu}_{\pi Q}(x)\,\epsilon_\nu.
\end{eqnarray}
$\hat{T}^{\mu\nu}_{\pi Q}(x)$ can be interpreted as a pseudo-local pion-quark EMT formally similar to the purely local one but with the quark source operator, $\vec{\bfvec{S}}(x)= \bar{q}(x)\bfvec{\gamma}\gamma_5 \vec{\tau} q(x)$, replaced by $\tilde{\bfvec{S}}(x)=\int d^3y\,\tilde{P}(\bfvec{x}-\bfvec{y})\,\vec{\bfvec{S}}(y)$. More generally, one can define an apparently local total EMT from the local one (Eq. \eqref{VIRIAL0}) by just systematically making this replacement, thereby incorporating the effect of the finite size of the pion. In particular, $\hat{T}^{00}$ exactly coincides with the Hamiltonian density introduced before; see Eq. \eqref{HTOT}. The non-conservation of this EMT tensor concerns only the spatial components, since the confining potential is static and the full Hamiltonian, which coincides with the volume integral of $\hat{T}^{00}$, is conserved. \\

Using the equation of motion for the quark field (Eqs. \eqref{HF1}, \eqref{HF2}, \eqref{REDLAG}, \eqref{QEOS}), one has:
\begin{eqnarray}
 \hat{T}^{00}(x)&=&\left(\mathcal{H}_{QK}\,+\,\mathcal{H}_{C}\,+\,\mathcal{H}_{\pi Q}\,+\,\mathcal{H}_{\pi c} 
\,+\,\mathcal{H}_{\pi K}\,+\,\mathcal{H}_{\pi M}\right)(x),\\
\hat{P}(x)=\frac{1}{3}\hat{T}^{ii}(x)&=&\left(\frac{1}{3}\mathcal{H}_{QK}\,+\,\frac{1}{3}\mathcal{H}_{\pi Q}\,+\,\mathcal{H}_{\pi c} 
\,-\,\frac{1}{3}\mathcal{H}_{\pi K}\,-\,\mathcal{H}_{\pi M}\right)(x),
\end{eqnarray}   
where the various Hamiltonian density operators are given by Eqs. \eqref{HTOT},\eqref{HDENS0},\eqref{HDENS},\eqref{HQ},\eqref{HPIQ},\eqref{HPI}, with $\vec{\bfvec{S}}(\bfvec{x})$ replaced by $\tilde{\bfvec{S}}(\bfvec{x})$. The operator $\hat{P}(x)$ can be referred to as the local pressure operator, in the sense that the average mechanical pressure is $\bar{P}=\int d^3 r\,\langle N|\hat{P}(\bfvec{r})|N\rangle /V$.
\subsection{The von Laue stability condition}
For any (normalized) stationary state $|\Psi\rangle$, one has, as a consequence of Eq.~\eqref{VIRIAL0},
\begin{eqnarray}
 \langle\Psi| \partial_i \hat{T}^{i k}(\bfvec{r}) |\Psi\rangle &=& \langle\Psi|\partial_\mu \hat{T}^{\mu k}(\bfvec{r}) |\Psi\rangle=-\langle\Psi|\partial_k W(\bfvec{r})\,q^\dagger(\bfvec{r})\, q(\bfvec{r})|\Psi\rangle,
\end{eqnarray}
where, again,  as in the rest of this discussion, the space-time point $x=(0,\bfvec{r}$) is implicitly taken at $t=0$. It follows that:
\begin{eqnarray}
  \int d^3r \,r^k \,\langle\Psi| \partial_i \hat{T}^{i k}(\bfvec{r}) |\Psi\rangle &=& -\int d^3r \,\langle\Psi|  \hat{T}^{i i}(\bfvec{r}) |\Psi\rangle\nonumber\\
 &=& -\int d^3r \,\langle\Psi|q^\dagger(\bfvec{r})\, \bfvec{r}\cdot \bfvec{\nabla}_{\bfvec{r}} W_C(\bfvec{r})\,q(\bfvec{r}) |\Psi\rangle.
\end{eqnarray}
Hence we establish  the virial theorem, which states:
\begin{equation}
 \int d^3r \,\langle\Psi|  \hat{T}^{i i}(\bfvec{r}) |\Psi\rangle   -\int d^3r \,\langle\Psi|q^\dagger(\bfvec{r})\, \bfvec{r}\cdot \bfvec{\nabla}_{\bfvec{r}} W_C(\bfvec{r})\,q(\bfvec{r}) |\Psi\rangle=0.
\end{equation}
Here, the potential $W_C$ can be viewed as an external potential acting on each of the constituent quarks that make up the nucleon core. However, this confining interaction arises from the existence of a three-branch string junction, which is at the origin of the very existence of the nucleon. Accordingly, we can consider that the local pressure operator should  be modified to incorporate the  presence of the confining interaction:
\begin{equation}
\hat{P}(x)=\frac{1}{3} \, \hat{T}^{i i}(x) \quad\to\quad \hat{P}(x)=\frac{1}{3} \, \hat{T}^{i i}(x)\,-\frac{1}{3} \,\,q^\dagger(x)\, \bfvec{x}\cdot \bfvec{\nabla}_{\bfvec{x}} W_C(\bfvec{x})\,q(x).
\end{equation}
The validity of this prescription is further reinforced by the possibility of defining  a confining stress operator, whose divergence supplies the confining force, namely,
$\hat{T}_C^{i k}(x)=-(1/2) \,\mathcal{P}_C(x)\left(\delta^{ik}\,-\,(r^i r^k/r^2)\right)$, with $\mathcal{P}_C(x)$ defined just below, or, said differently, its divergence just compensates that of $\hat{T}^{i k}(\bfvec{r})$
according to Eq. (66) (see appendix A for more details). Even though this definition of the confinement pressure may require further justification, it nevertheless allows us to gain, at least at an exploratory level, an understanding of the distribution of forces within the nucleon.\\
We recall that, in Eq. \ref{eqCPP}, we have computed a useful quantity called ``confining pressure potential" to calculate $\hat P$.\\
If we now introduce the following notations: 
\begin{eqnarray}
\mathcal{P}_C(x)&=& \,q^\dagger(x)\, \bfvec{x}\cdot \bfvec{\nabla}_{\bfvec{x}}W_C(\bfvec{x})\,q(x)\equiv \hat{P}_C(x)\label{PCONFF}\\ 
H'_C &=&\int d^3r \,q^\dagger(\bfvec{r})\, \bfvec{r}\cdot \bfvec{\nabla}_{\bfvec{r}}W_C(\bfvec{r})\,q(\bfvec{r})\\
P_C &=&\langle N|H'_C|N\rangle,
\end{eqnarray}
then the physically motivated, modified local  pressure  operator finally reads:
\begin{equation}
\hat{P}(x)=\left(\frac{1}{3}\mathcal{H}_{QK}\,-\,\frac{1}{3} \,\mathcal{P}_C\,+\,\frac{1}{3}\mathcal{H}_{\pi Q}\,+\,\mathcal{H}_{\pi c} 
\,-\,\frac{1}{3}\,\mathcal{H}_{\pi K}\,-\,\mathcal{H}_{\pi M}\right)(x).
\end{equation}
The virial theorem for the bound nucleon, $\int d^3r\,\langle \hat{P}\rangle(\bfvec{r})=0$, takes the explicit form with a transparent notation for the energy $E_A$ referring to the Hamiltonian piece ${H}_A$ (see Eqs. \eqref{HTOT},\eqref{HDENS0},\eqref{HDENS},\eqref{HQ},\eqref{HPIQ}, and \eqref{HPI}),
\begin{eqnarray}
&&\frac{1}{3}\, E_{QK}\,-\, \frac{1}{3}\,P_C\,+\,\frac{1}{3}\, E_{\pi Q} \,+\, 
E_{\pi c} \,-\,\frac{1}{3}\,E_{\pi K} \,-\,E_{\pi M} =0
\label{VONLAUE},
\end{eqnarray}
which is the venerable von Laue stability condition \cite{vonLaue}, whereas the total energy of the bound nucleon is:
\begin{eqnarray}
E_0&=&E_{QK}\,+\,E_{QM}\,+ E_C\,+\, E_{\pi Q} \,+\, 
E_{\pi c} \,+\,E_{\pi K} \,+E_{\pi M}. 
\end{eqnarray}
The physical meaning is very transparent: the quark Fermi pressure, $E_{QK}/3$, is balanced by the negative bag pressure, $- P_C/3,$ and the negative pion pressure, which in practice is very close to the pion cloud kinetic pressure $-P_{\pi}\sim - E_{\pi K}$. \\


Eq.\eqref{VONLAUE} is an exact result, and this von Laue condition should be satisfied when using the Fock-space amplitude $\Phi\left({\bf p}_1, {\bf p}_2, {\bf p}_3\right)$, solution of the HF Dirac equation. However, in practice, and for the purposes of this paper and the companion one, it will serve as an alternative to the variational principle used in non-relativistic quantum mechanics. In practice, the three-quark Fock-space amplitude  will always  depend on the Gaussian single-quark Fock-space amplitude :
\begin{equation}
\Phi(\bfvec{p})= R_0(p)= (2\pi)^{\frac{3}{2}}\,\left(\frac{b^2}{\pi}\right)^{\frac{3}{4}}\,e^{-b^2\,p^2/2}. 
\label{TRIALWF}\end{equation}
Throughout the remainder of this paper, the value of the size parameter $b$ will thus be fixed by requiring that the von Laue stability condition is satisfied, and the determination of the ``best value" of the quark core size parameter $b$ is the central issue of the paper.
\section{Static EMT tensor: Energy density and pressure distribution}
\subsection{The static EMT tensor}
The static EMT tensor $t^{\mu\nu}(\bfvec{r})$ is defined  from the matrix element evaluated in the Breit frame of the  EMT tensor according to \cite{Polyakovor}, and we also recall that this static EMT represents the matrix element of $\hat{T}^{\mu\nu}(0,\bfvec{r})$ in the Wigner sense, as defined in Eq. \ref{WIGDEF}:
\begin{eqnarray}
t^{\mu\nu}(\bfvec{r})&=&\int\frac{d^3\Delta}{(2\pi)^3}\, e^{-i\,\bfvec{\Delta}\cdot\bfvec{r}} \,\left\langle\frac{\bfvec{\Delta}}{2}\left| \,\hat{T}^{\mu\nu}(0,\bfvec{0})\,\right|-\frac{\bfvec{\Delta}}{2}\right\rangle=\int\frac{V d^3\Delta}{(2\pi)^3}
\left\langle\frac{\bfvec{\Delta}_N}{2}\left|\, \hat{T}^{\mu\nu}(0,\bfvec{r})\,\right|-\frac{\bfvec{\Delta}_N}{2}\right\rangle .
\end{eqnarray}
To account for the effect of confining pressure, we also introduce a Wigner density of the ``confining pressure potential''  that may be viewed as a ``static confining pressure", defined as follows:
\begin{eqnarray}
p_C(\bfvec{r})&=&\int\frac{d^3\Delta}{(2\pi)^3}\, e^{-i\,\bfvec{\Delta}\cdot\bfvec{r}} \,\left\langle\frac{\bfvec{\Delta}}{2}\left| \,\hat{P}_C(0,\bfvec{0})\,\right|-\frac{\bfvec{\Delta}}{2}\right\rangle=\int\frac{V d^3\Delta}{(2\pi)^3}
\left\langle\frac{\bfvec{\Delta}_N}{2}\left|\, \hat{P}_C(0,\bfvec{r})\,\right|-\frac{\bfvec{\Delta}_N}{2}\right\rangle .
\end{eqnarray}
At variance with the definition given in \cite{Goeke,Polyakov,Lorce2019,Lorce2021}, where a covariant normalization is used, the momentum states here are normalized either as in non relativistic quantum mechanics, $\left\langle \bfvec{P}' |\bfvec{P}\right\rangle=(2\pi)^3\delta^{(3)}(\bfvec{P}'-\bfvec{P})$, or to one in a large box of volume $V$, ($\left\langle \bfvec{P}'_N |\bfvec{P}_N\right\rangle=\delta_{\bfvec{P}'_N,\bfvec{P}_N})$. In the following, we will omit the reference to the arbitrary time $t=0$, since in the Breit frame the two nucleon states, $|-\bfvec{\Delta}/2\rangle$ and $|\bfvec{\Delta}/2\rangle$, have the same energy. Owing to translational invariance, the matrix element of the EMT component can be identically expressed as: 
\begin{equation}
\left\langle\frac{\bfvec{\Delta}}{2}\left| \,\hat{T}_Y(0,\bfvec{0})\,\right|-\frac{\bfvec{\Delta}}{2}\right\rangle  =\int\frac{d^3 y}{V}  \,e^{i\,\bfvec{\Delta}\cdot\bfvec{y}} \,\left\langle\frac{\bfvec{\Delta}}{2}\left| \,\hat{T}_Y(o,\bfvec{y})\,\right|-\frac{\bfvec{\Delta}}{2}\right\rangle .  
\end{equation}
By construction, the volume integral of the above Wigner density reproduces the expectation value of the corresponding component of  the EMT tensor:
\begin{equation}
 \int d^3r\, t^{\mu\nu}(\bfvec{r})= \int d^3y\, \langle  0_P|\hat{T}^{\mu\nu}(0,\bfvec{y})|0_P\rangle.
\end{equation}
\subsection{Energy density and pressure distribution}
The various components of the EMT tensor are observable quantities in the sense that they are experimentally directly obtained from the generalized parton distributions (GPDs) \cite{Polyakov,Lorce2019}, which are accessible in several exclusive processes, such as deeply virtual Compton scattering \cite{DVCS} and meson production \cite{MESPROD}. It has been proposed \cite{Polyakov} that the components  $t^{00}$ and the component $t^{ii}$ can be related to the genuine energy density and the genuine mechanical isotropic pressure distribution  inside the nucleon according to:
\begin{eqnarray}
\varepsilon(r)= t^{00}(\bfvec{r}), \ \qquad &\hbox{with:}&\qquad  M_N=\int 4\pi r^2\,dr\,\varepsilon(r) ,\\
p(r)=\frac{1}{3}t^{ii}(\bfvec{r})-\frac{1}{3}p_C(\bfvec{r}),  \qquad &\hbox{with  the von Laue condition: }&\qquad  
\int 4\pi r^2\,dr\,p(r)=0.
\end{eqnarray}
The explicit decompositions of the energy density and the pressure distribution are: 
\begin{eqnarray}
 \varepsilon(r)&=&\varepsilon_{QK}(r)\,+\,\varepsilon_{QM}(r)\,+ \varepsilon_C(r)\,+\, \varepsilon_{\pi Q}(r) \,+\, 
\varepsilon_{\pi c}(r) \,+\,\varepsilon_{\pi K}(r) \,+\varepsilon_{\pi M}(r)\ ,\\  
p(r)&=&\frac{1}{3}\varepsilon_{QK}(r)\,-\frac{1}{3} p_C(r)\,+\,\frac{1}{3} \varepsilon_{\pi Q}(r) \,+\, 
\varepsilon_{\pi c}(r) \,-\,\frac{1}{3}\varepsilon_{\pi K}(r) \,-\varepsilon_{\pi M}(r).
\end{eqnarray}
In an analogous way, and as discussed and justified in the introduction, one can associate with each local quark density operator of the type, $\hat{T}_\Gamma(\bfvec{r})=\bar{q}(\bfvec{r})\,\Gamma \,q(\bfvec{r})$, a density, i.e., a spatial distribution  in the Wigner sense according to:
\begin{eqnarray}
  t_\Gamma (\bfvec{r})=
\int \frac{V d^3\Delta}{(2\pi)^3}\, \left\langle \bfvec{\Delta}_N/2\right|\hat{T}_\Gamma (\bfvec{r}) \left| -\bfvec{\Delta}_N/2\right\rangle=\int \frac{V d^3\Delta}{(2\pi)^3}\,e^{-i\bfvec{\Delta}\cdot\bfvec{r}} \left\langle \bfvec{\Delta}_N/2\right|\hat{T}_\Gamma (0) \left| -\bfvec{\Delta}_N/2\right\rangle.
\end{eqnarray}
\section{Localized nucleon state: Static nucleonic bag }
\subsection{Quark kinetic, mass  and confining contributions to the energy and pressure for the static nucleonic bag}
In this section, we specialize in the case of a localized nucleon state, $|N: \bfvec{0}_X\rangle\,\equiv\,|N(\bfvec{X}_N=0)\rangle$, for which the position of the center of mass coincides with the origin of the coordinate system, which itself coincides in the CCM with the string junction position. We refer to such a system as the ``static nucleonic bag." The Fock-space amplitude is just the product of three identical single-quark amplitudes, $\Phi\left({\bf p}_1, {\bf p}_2, {\bf p}_3\right)=\Phi({\bf p}_1)\Phi( {\bf p}_2)\Phi({\bf p}_3)$, (properly projected onto the appropriate spin-flavor-color singlet state, see Eq.~\eqref{WFNOCM}). Each component of the  quark energy, as well as the average quark confining pressure $\bar P$,  can be written as a volume integral of a quark energy density and a pressure distribution, according to:
\begin{eqnarray}
 E_Q &=& E_{QK}+ E_{QM}+E_C= \int d^3 r\,\varepsilon_Q(r)= \int d^3 r\,\varepsilon_{QK}(r)\,+\,\int d^3 r\,\varepsilon_{QM}(r)\,+\,\int d^3 r\,\varepsilon_C(r)\ ,  \label{EQTOT}\nonumber\\
\bar{P}\,V &=&\frac{1}{3} E_{QK}-\frac{1}{3}P_C=\frac{1}{3} \int d^3 r\,\varepsilon_{QK}(r)\,-\,\frac{1}{3}\int d^3 r\,\mathcal{P}_C(r),\nonumber\\
 \varepsilon_{QK}(r)&=&\left\langle N:\bfvec{0}_X\right|q^\dagger(\bfvec{r})\left[-i\bfvec{\alpha}\cdot\vec\nabla_{\bfvec{r}} \right]\,q(\bfvec{r}) \left|N:\bfvec{0}_X\right\rangle \nonumber\\
 &=&3\,\left( u(r)v'(r)-u'(r) v(r)\,+\,\frac{2 u(r) v(r)}{r}\right)\ ,\label{EPSQK}\\
 \varepsilon_{QM}(r)&=&\left\langle N:\bfvec{0}_X\right|q^\dagger(\bfvec{r}) \beta\,M_q\,q(\bfvec{r}) \left|N:\bfvec{0}_X\right\rangle=3\,M_q\,\left(u^2(r) - v^2(r)\right)\ ,\label{EPSQM}\\
 \varepsilon_{C}(r)&=&\left\langle N:\bfvec{0}_X\right|q^\dagger(\bfvec{r})\,W_C(\bfvec{r})\,q(\bfvec{r}) \left|N:\bfvec{0}_X\right\rangle = 3\,W_C(r)\,\left(u^2(r) +v^2(r)\right)\  , \label{EPSQC}\\
 -\mathcal{P}_C(r)&=&-\left\langle N:\bfvec{0}_X\right|q^\dagger(\bfvec{r})\,\bfvec{\nabla}_{\bfvec{r}}W_C(\bfvec{r})\,q(\bfvec{r}) \left|N:\bfvec{0}_X\right\rangle=
 -3\,r\,W'_C(r)\,\left(u^2(r) +v^2(r)\right) \label{EPSQCC}.
\end{eqnarray}
Here $u(r)$ and $v(r)$ are the up and down components of the QPW wave function, as given in \eqref{QPW2}, \eqref{QPW1}, \eqref{QPW3}, \eqref{QPW4}. 
Alternatively, the quark kinetic and mass energies can be calculated more directly in momentum space:
\begin{eqnarray}
E_{QK} &=&3\int \frac{d^3p}{(2\pi)^3}\,\frac{p^2}{E_p}\,\Phi^2(p),\\
E_{QM} &=&3\int \frac{d^3p}{(2\pi)^3}\,\frac{M_q^2}{E_p}\,\Phi^2(p),\\
E_{QK}+E_{QM}&=&3\int \frac{d^3p}{(2\pi)^3}\,E_p\,\Phi^2(p).
\end{eqnarray}
\subsection{Pure confining potential}
Let us first consider the case in which the pion cloud is ignored, that is, the case in which the quarks experience only the confining potential. The Hartree-Fock equation reduces to:
\begin{equation}
\frac{\delta\left(\langle|N(\Phi)|H|N(\Phi)\rangle\,-\,3\epsilon_0\, \int \frac{d^3 p}{(2\pi)^3}\,|\left|\Phi({\bf p})\right|^2\right)}{\delta \Phi^*({\bf p})} =0.     
\end{equation}
The total HF energy, which is identical to the total energy, is just three times the single-particle energy, i.e., $E_{HF}=E_0=3\epsilon_0$. The explicit form of this HF equation is formally similar to a Schr\"odinger-like bound state equation for the two component spinor: 
$$\hat{\Phi}_s(\bfvec{p})=\Phi(\bfvec{p})\,\chi_s=R_0(p)\,\chi_s .$$
This bound state equation has already been derived in our previous papers using a different method \cite{Universe,Universe2025}. With the present notations, this eigenvalue equation reads:
\begin{eqnarray}
&&E_p\,\hat{\Phi}(\bfvec{p})\,+\,\frac{1}{2}\,\int\frac{d^3 q}{(2\pi)^3}\,\tilde{W}_C({\bf p} - {\bf q})\,\bigg(\sqrt{1+s_p}\sqrt{1+s_q}\nonumber\\
&&+\sqrt{1-s_p}\sqrt{1-s_q}\,\bfvec{\sigma}\cdot {\bf p}\,\bfvec{\sigma}\cdot {\bf q}\bigg)\,\hat{\Phi}(\bfvec{p})=\epsilon_0\,\hat{\Phi}(\bfvec{q}),\label{HFEQ}
\end{eqnarray}
where $\tilde{W}_C({\bf k})=\int d^3r\,e^{-i\bfvec{k}\cdot\bfvec{r}}_, W_C(\bfvec{r})$ is the Fourier transform of the confining interaction. The bound-state HF energy can be written as (see Eqs. \eqref{EQTOT}, \eqref{EPSQK}, \eqref{EPSQM}), \eqref{EPSQC}:
\begin{equation}
\epsilon_0= \int d^3 x\, \left(\psi\right)^\dagger(\bfvec{x})\left[-i\bfvec{\alpha}\cdot\vec\nabla_{\bfvec{x}}\,+\, \beta M_q \, +  \,\,W_C(\bfvec{x})\right]\,\left(\psi\right)(\bfvec{x}),   
\end{equation}
where $\psi(\bfvec{x})$ is the quasi plane-wave (QPW) wave function:
\begin{eqnarray}
 \left(\psi\right)({\bf{r}})&=& \int\frac{d^3 p}{(2\pi)^3} \,  e^{-i {\bf p}\cdot {\bf r}}\,
\sqrt{\frac{E_P + \mathcal{S}}{2 E_P}}\left(
\begin{array}{c}
	\hat{\Phi}_s({\bf{p}})\\
	\frac{\bfvec{\sigma}\cdot {\bf p}}{E_P + \mathcal{S}}\,\,\hat{\Phi}_s({\bf{p}})
\end{array}\right)\equiv \left(
\begin{array}{c}
	u(r)\,\chi_s\\
	i\,\bfvec{\sigma}\cdot \hat{\bf r}\,\,v(r)\,\chi_s
\end{array}\right). 
\end{eqnarray}

Hence, the problem of determining the bound state constructed on a BCS-like vacuum (Eqs. \eqref{NUCWF}, \eqref{WFNOCM}) is formally equivalent to solving a Dirac equation for a quark moving in an effective central potential $W_C(r)$, but with solutions limited to the subset of QPW wave functions. 
One can also view the problem as a variational approach to a relativistic quantum mechanical problem for a particle moving in an external confining potential, with the variational space restricted to QPWs. If, in addition, one restricts to a subset in which the Fock-space amplitude is Gaussian (eq. \eqref{TRIALWF}), the question then becomes that of determining the optimal value of the parameter 
$b$. Now the key point is: what criterion should be used to find the optimal trial Fock-space amplitude, that is, what is the optimal value of the parameter $b$ controlling the size of the trial Gaussian wave function? \\

\begin{figure}
\includegraphics[width=0.8\textwidth,angle=0]{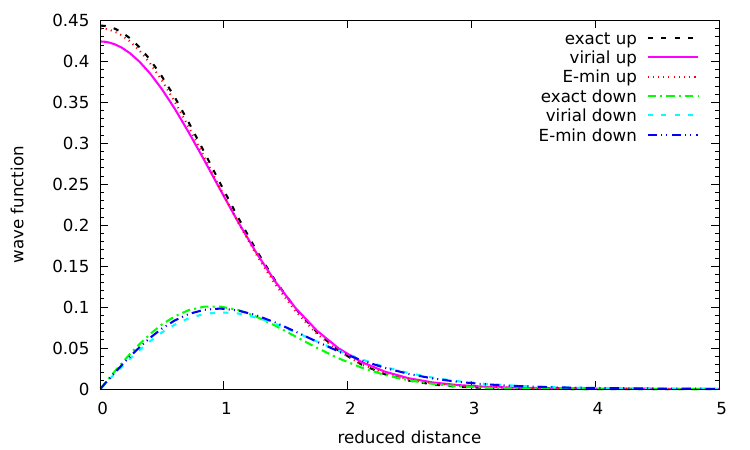}
\caption{The up and down Dirac wave functions in an equally mixed scalar-vector relativistic OH  potential versus the reduced distance $\sqrt{\sigma}r$. ``exact" stands for the exact result. ``virial" stands for the solution obtained with a trial Gaussian QPW wave function satisfying the von Laue stability condition. ``E-min" stands for the case with a trial Gaussian QPW  wave function that minimizes the energy.}
\label{QWF}
\end{figure}

To gain some insight into this question, we consider the relativistic quantum mechanical problem of a particle with mass $M_q$ moving in a relativistic harmonic-oscillator potential,

\begin{equation}
 W_{RHO}(r)=  \frac{1}{2}\left(1\,+\,\gamma^0\right)\,\sigma^{3/2} \,r^2,
\end{equation}
for which an exact analytical solution exists. In particular, the ground state  energy eigenvalue satisfies:
$$\epsilon_{RHO}=M_q\,+\,\sqrt{\frac{2 \sigma^{3/2}}{\epsilon_{RHO} + M_q}}.$$
If we take for orientation the NJLset1 parameters of \cite{Universe2025}, namely, $\sigma=0.18\,\rm{\rm{GeV}}^2$, $M_q=356.7\,\rm{MeV}=0.841\,\sqrt{\sigma}$, we find $\epsilon_{RHO}=2.082\,\sqrt{\sigma}$. We can  verify numerically that the virial theorem \eqref{VONLAUE},  
\begin{equation}
\int d^3 r\, q^\dagger(\bfvec{r})\left[-i\bfvec{\alpha}\cdot\bfvec{\nabla}_{\bfvec{r}}\right]\,q(\bfvec{r}) - \int d^3r \,q^\dagger(\bfvec{r})\, \bfvec{r}\cdot \bfvec{\nabla}_{\bfvec{r}}W_C(\bfvec{r})\,q(\bfvec{r})=0,
\end{equation}
also equivalent in this particular case to  $E_{QK}=2 E_C$, is exactly satisfied (see also the discussion in Appendix A of Ref. \cite{Lorce2021}). Regarding the trial Gaussian QPW wave function, the minimal value of the orbital energy, $\epsilon_{min}= 2.060\,\sqrt{\sigma}$, is obtained for $b_{min}=0.878/\sqrt{\sigma}$. We observe that, indeed, this energy, being lower than the exact energy, does not constitute an upper bound. The value of the size parameter for which the virial theorem (i.e., the von Laue condition) is satisfied is slightly larger, $b_{vir}=0.91/\sqrt{\sigma}$, and the value of the orbital energy is  $\epsilon_{vir}= 2.061\,\sqrt{\sigma}$. These results do not allow us to conclude whether the QPW solution satisfying the virial theorem corresponds to the Hartree-Fock solution of Eq. \eqref{HFEQ}. However, as a side remark, if one is interested in the approximate solution of the relativistic harmonic oscillator problem that satisfies the virial theorem-i.e., one that favors the condition of mechanical stability- this solution yields an energy value slightly closer to the exact one than the solution that minimizes the Hartree-Fock energy. 
This energy associated with the trial Gaussian QPW wave function reproduces the exact result with good accuracy, underestimating it by only $1\%$. In Fig. \ref{QWF} we display the up and down quark wave functions, $u(r)$ and $v(r)$, for these three cases. 
If we now return to the case of the chiral confining model with the same parameters: we find: $b_{min}=0.976/\sqrt{\sigma}=0.454\,\rm{fm}$,  $\epsilon_{min}=1.418\,\sqrt{\sigma}=0.601\,\rm{GeV}$, $b_{vir}=0.933/\sqrt{\sigma}=0.434\,\rm{fm}$,  $\epsilon_{vir}=1.421\,\sqrt{\sigma}=0.602\,\rm{GeV} $. \\

From this brief but instructive study, one can draw conclusions about an appropriate way to implement the virial theorem, since the question of the overall mechanical stability of the nucleon is our main goal. First, one can observe that the HF solution, which belongs to the subset of QPW wave functions, has no reason, in general, to satisfy the virial theorem, since, as is explicitly illustrated by the example of the relativistic harmonic oscillator, the virial theorem is satisfied by the exact solution. This is why it may be advantageous to consider this problem using a variational approach, with the variational space restricted to QPWs. The often-used energy minimization, as in Ref. \cite{Pirner}, is, however, sometimes questionable because, due to the existence of negative-energy solutions, the minimum energy is not necessarily an upper bound. This is indeed the case in the example of the relativistic harmonic oscillator, for which the minimum value of the energy is lower than the exact value, which is self-contradictory with the underlying principle of the method; it also leads to a solution which has no reason to ensure global mechanical equilibrium. By contrast, if the virial theorem is used as an alternative variational principle, there is nothing that prevents the resulting energy from being slightly lower than the exact energy. Taken together, these considerations lead us to favor the von Laue stability condition, since satisfying the virial theorem is fundamentally a statement about mechanical equilibrium, which is our main focus in this paper and the companion paper. Hence, this von Laue condition will be systematically implemented in what follows.

\subsection{The pion cloud contribution to the energy in the static CCM nucleonic bag}\label{PIENERGY}
The energies associated with the pionic sector of the Hamiltonian—namely the quark–pion interaction term, $H_{\pi Q}$ (Eq. \eqref{HPIQ}), and the purely pionic Hamiltonian, $H_{\pi\pi}=H_{\pi c}\,+\,H_{\pi K}\,+\,H_{\pi M}$ \eqref{HPI}, which  consists of contact, kinetic, and mass contributions—have to be evaluated. In addition, all individual components, $H_{\pi Q},\ H_{\pi c}, H_{\pi K}, H_{\pi M}$, must be considered  separately in order to formulate the von Laue stability condition. These quantities can be evaluated using Green’s function techniques.

To this end, we first expand the pion field and its conjugate momentum following the standard mode decomposition introduced earlier (Eqs. \eqref{PIONFIELD}, \eqref{PIONCONJ}). Applying Green’s function methods, as developed for example in Refs. \cite{CSN1990,Chanfray2007},  the different contributions to the pion cloud energy can then be derived as follows:
\begin{eqnarray}
 E_{\pi M}&=& \left\langle N :\bfvec{0}_X\right| H_{\pi M}\left|N :\bfvec{0}_X\right\rangle=\frac{M_\pi^2}{2}\,\int d^3 r \,\left\langle N:\bfvec{0}_X\right| \vec{\Phi}^2(\bfvec{r})\left|N:\bfvec{0}_X\right\rangle\nonumber\\
 &=&\frac{3}{2}\sum_{\bfvec{k}}\, \frac{M_\pi^2}{2\,\omega_k}\,i\, \lim_{\,t'\to t^+}\,G^{\Phi\Phi}_{\bfvec{k}}(t,t')\nonumber\\
 &\equiv& \frac{3}{2}\sum_{\bfvec{k}}\, \frac{M_\pi^2}{2\,\omega_k}\,i\, \lim_{\,t'\to t^+}\left[-i\left\langle N:\bfvec{0}_X\right| \mathcal{T}\left((b_{\bfvec{k}a}\,+\,b^\dagger_{-\bfvec{k} a})(t)(b^\dagger_{\bfvec{k}a}\,+\,b_{-\bfvec{k} a})(t')\right)\left|N:\bfvec{0}_X\right\rangle\right]\nonumber\\
 &=&\frac{3}{2}\sum_{\bfvec{k}}\, \frac{M_\pi^2}{2\,\omega_k}\,\int\frac{id\omega}{2\pi}\,e^{i\omega\eta^+}\,\frac{2\omega_k}{\omega^2-\omega_k^2-k^2\Pi_0(k,\omega)/(4F^2_\pi V )}\nonumber\\
 &=&\frac{3}{2}\,\left(\frac{1}{2F_\pi}\right)^2\,\int\frac{d^3 k}{(2\pi)^3}\,\frac{k^2 M_\pi^2}{2\,\omega_k}\int\frac{id\omega}{2\pi}\,e^{i\omega\eta^+}\,
 \frac{2\,\omega_k \Pi_0(k,\omega)}{(\omega^2-\omega_k^2 + i\eta)^2}\label{EPIM} 
\end{eqnarray}
\begin{eqnarray}
 E_{\pi K}&=& \left\langle N:\bfvec{0}_X \right| H_{\pi K}\left|N:\bfvec{0}_X\right\rangle=\frac{1}{2}\,\int d^3 r \,\left\langle N:\bfvec{0}_X\right| (\bfvec{\nabla}\vec{\Phi})^2(\bfvec{r})\left| N:\bfvec{0}_X\right\rangle\nonumber\\
 &=&\frac{3}{2}\sum_{\bfvec{k}}\, \frac{k^2}{2\,\omega_k}\,i\, \lim_{\,t'\to t^+}\,G^{\Phi\Phi}_{\bfvec{k}}(t,t')\nonumber\\
 &\equiv& \frac{3}{2}\sum_{\bfvec{k}}\, \frac{k^2}{2\,\omega_k}\,i\, \lim_{\,t'\to t^+}\left[-i\left\langle N:\bfvec{0}_X\right| \mathcal{T}\left((b_{\bfvec{k}a}\,+\,b^\dagger_{-\bfvec{k} a})(t)(b^\dagger_{\bfvec{k}a}\,+\,b_{-\bfvec{k} a})(t')\right)\left|N:\bfvec{0}_X\right\rangle\right]\nonumber\\
 &=&\frac{3}{2}\sum_{\bfvec{k}}\, \frac{k^2}{2\,\omega_k}\,\int\frac{id\omega}{2\pi}\,e^{i\omega\eta^+}\,\frac{2\omega_k}{\omega^2-\omega_k^2-k^2\Pi_0(k,\omega)/(4F^2_\pi V)}\nonumber\\
 &=&\frac{3}{2}\,\left(\frac{1}{2F_\pi}\right)^2\,\int\frac{d^3 k}{(2\pi)^3}\,\frac{k^4}{2\,\omega_k}\int\frac{id\omega}{2\pi}\,e^{i\omega\eta^+}\,
 \frac{2\,\omega_k \Pi_0(k,\omega)}{(\omega^2-\omega_k^2 + i\eta)^2}\label{EPIK} 
\end{eqnarray}
\begin{eqnarray}
 E_{\pi c}&=& \left\langle N:\bfvec{0}_X\right| H_{\pi c}\left|N:\bfvec{0}_X\right\rangle=\frac{1}{2}\,\int d^3 r \,\left\langle N:\bfvec{0}_X\right|\vec{\Pi}^2(\bfvec{r}) \left|N:\bfvec{0}_X\right\rangle\nonumber\\ 
 &=&\frac{3}{2}\sum_{\bfvec{k}}\, \frac{\omega^2_k}{2\,\omega_k}\,i\, \lim_{\,t'\to t^+}\,G^{\Pi\Pi}_{\bfvec{k}}(t,t')\nonumber\\
 &\equiv& \frac{3}{2}\sum_{\bfvec{k}}\, \frac{\omega^2_k}{2\,\omega_k}\,i\, \lim_{\,t'\to t^+}\left[-i\left\langle N:\bfvec{0}_X\right| \mathcal{T}\left((b_{\bfvec{k}a}\,-\,b^\dagger_{-\bfvec{k} a})(t)(b^\dagger_{\bfvec{k}a}\,-\,b_{-\bfvec{k} a})(t')\right)\left|N:\bfvec{0}_X\right\rangle\right]\nonumber\\
 &=&\frac{3}{2}\sum_{\bfvec{k}}\, \frac{\omega^2_k}{2\,\omega_k}\,\int\frac{id\omega}{2\pi}\,e^{i\omega\eta^+}\,\frac{2\omega_k + 2 k^2\Pi_0(k,\omega)/(8\omega_k F^2_\pi)} {\omega^2-\omega_k^2-k^2\Pi_0(k,\omega)/(4F^2_\pi V)}\nonumber\\
 &=&\frac{3}{2}\left(\frac{1}{2F_\pi}\right)^2\,\int\frac{d^3 k}{(2\pi)^3}\frac{k^2 \omega^2k}{2\,\omega_k}\int\frac{id\omega}{2\pi}e^{i\omega\eta^+}\,
\bigg[ \frac{2\omega_k \Pi_0(k,\omega)}{(\omega^2-\omega_k^2 + i\eta)^2} +\frac{\Pi_0(k,\omega)/\omega_k}{\omega^2-\omega_k^2 + i\eta}\bigg],\label{EPIC}
\end{eqnarray}
with $\omega_k=\sqrt{k^2+M^2_\pi}$. In the last line of the three  expressions above, the free pion contribution has been subtracted. $\Pi_0(k,w)$ denotes the Fourier transform of a polarization propagator according to: 
\begin{eqnarray}
\Pi_0(k; t,t')&=&  -i\left\langle N:\bfvec{0}_X\right| \mathcal{T}\left(\tilde{S}^\dagger_{\bf k}(t), \tilde{S}_{\bf k}(t')\right)\left|N:\bfvec{0}_X\right\rangle  = \int\frac{d\omega}{2\pi}\,e^{-i\,\omega\,(t-t')}\,\Pi_0(k,w)\\
\Pi_0(k,w)&=&\sum_n\,\left(\frac{|\left\langle n\right|\tilde{S}_{\bf k} \left|N:\bfvec{0}_X\right\rangle|^2}{\omega-E_n+i\eta}-
\frac{|\left\langle n\right|\tilde{S}^\dagger_{\bf k} \left|N:\bfvec{0}_X\right\rangle|^2}{\omega+E_n -i\eta}\right). 
\end{eqnarray}
The equation of motion for the pion field is:
\begin{eqnarray}
\frac{\partial^2 }{\partial t^2}\vec{\Phi}(\bfvec{r}, t)&=&-\big[H,\big[H,\vec{\Phi}(\bfvec{r}, t)\big]\big]=i\,\big[H,\vec{\Pi}(\bfvec{r}, t)\big]\\
&=& \nabla^2\vec{\Phi}(\bfvec{r},t)- M^2_\pi\,\vec{\Phi}(\bfvec{r},t)- \frac{1}{2 F_\pi}\,\bfvec{\nabla}\cdot\tilde{\bfvec{S}}(\bfvec{r},t).\label{EOMPI}
\end{eqnarray}
It follows that:
\begin{eqnarray}
-i\left\langle N:\bfvec{0}_X\right| \mathcal{T}\left((b_{\bfvec{k}a}\,+\,b^\dagger_{-\bfvec{k} a})(t), \tilde{S}_{\bfvec{k}a}(t')\right)\left|N:\bfvec{0}_X\right\rangle=\frac{1}{2 F_\pi} \int\frac{id\omega}{2\pi}\,e^{-i\omega\,(t-t')}
\sqrt{\frac{2\omega_k}{V}} \frac{k\,\Pi_0(k,\omega)}{\omega^2-\omega_k^2 + i\eta},   
\end{eqnarray}
and the pion-quark interaction energy also follows:
\begin{eqnarray}
E_{\pi Q}&=& \left\langle N:\bfvec{0}_X\right| H_{\pi Q}\left|N:\bfvec{0}_X\right\rangle =-\frac{1}{2 F_\pi} \,\int d^3r \left\langle N:\bfvec{0}_X\right|\tilde{\bfvec{S}}(\bfvec{r})\cdot \bfvec{\nabla}\vec{\Phi}(\bfvec{r}) \left|N:\bfvec{0}_X\right\rangle\nonumber\\
&=&-\frac{1}{2 F_\pi}\sum_{\bfvec{k}}\, \frac{k}{\sqrt{2\,\omega_k V}}i\, \lim_{\,t'\to t^+}\left[-i\left\langle N:\bfvec{0}_X\right| \mathcal{T}\left((b_{\bfvec{k}a}\,+\,b^\dagger_{-\bfvec{k} a})(t), \tilde{S}_{\bfvec{k}a}(t')\right)\left|N:\bfvec{0}_X\right\rangle\right]\nonumber\\
&=&3\,\left(\frac{1}{F_\pi}\right)^2\,\int\frac{d^3 k}{(2\pi)^3}\,\frac{k^2}{2\omega_k}\,\int\frac{id\omega}{2\pi}\,e^{i\omega\eta^+}\,
 \frac{2\omega_k\, \Pi_0(k,\omega)}{\omega^2-\omega_k^2 + i\eta}.\label{EPIQ} 
\end{eqnarray}
 The analytical structure of the integrand appearing in the last line of Eqs. \eqref{EPIM},\eqref{EPIK},\eqref{EPIC},\eqref{EPIQ}, considered as a function of $\omega$,  is such that it has a cut on the real axis and the corresponding propagators are analytic in the first
and third quadrants of the complex $\omega$-plane. In other words, the set of poles lies below the real axis for $\omega$ positive and above it for $\omega$ negative. For this reason, the (practical) calculations can be done using a Wick rotation. In practice, two integrals are  needed:
\begin{eqnarray}
 \int\frac{id\omega}{2\pi}\,\ e^{i\omega\eta^+}\,  \frac{ \Pi_0(k,\omega)}{\omega^2-\omega_k^2 + i\eta}&=&
 \int\frac{dz}{2\pi}\,  \frac{ \Pi_0(k,iz)}{z^2+\omega_k^2}=-\sum_n\,\frac{|\left\langle n\right|\tilde{S}_{\bf k} \left|N\right\rangle|^2}{\omega_k (\omega_k+E_n)}~,\\
 \int\frac{id\omega}{2\pi}\,\ e^{i\omega\eta^+}\,  \frac{ \Pi_0(k,\omega)}{(\omega^2-\omega_k^2 + i\eta)^2}&=&
 -\int\frac{dz}{2\pi}\,  \frac{ \Pi_0(k,iz)}{(z^2+\omega_k^2)^2}
 \nonumber\\
 &=&\sum_n\,\frac{1}{2\omega^2_k}\left(\frac{|\left\langle n\right|\tilde{S}_{\bf k} \left|N\right\rangle|^2}{\omega_k (\omega_k+E_n)}+\frac{|\left\langle n\right|\tilde{S}_{\bf k} \left|N\right\rangle|^2}{(\omega_k+E_n)^2}\right)~.
\end{eqnarray}
Concerning the summation over intermediate states $\left|n\right\rangle$, we limit ourselves to the nucleon states, $(I,J)_N=(1/2,1/2)$, and $\Delta(1232)$  states, $(I,J)_\Delta=(3/2,3/2)$, assuming identical orbital quark wave functions in both cases. The transition matrix element - that is, the transition form factor is written as:
\begin{eqnarray}
\left\langle n \bfvec{k}\right|\tilde{S}_{\bf k} \left|N:\bfvec{0}_X\right\rangle 
&\equiv& 
\left\langle \bfvec{0}_X :(I,J)_n \right|\,\tilde{S}_{\bf k}\, \left|\bfvec{0}_X\right\rangle= P (\bfvec{k}) \,\, \left\langle \bfvec{0}_X :(I,J)_n\right| \int d^3r \,e^{i{\bf k}\cdot{\bf r}}\,\bar{q}(\bfvec{r})\,\bfvec{\gamma}\cdot{\hat{\bf k}}\,\vec{\tau}\,q(\bfvec{r})\left|\bfvec{0}_X\right\rangle\nonumber\\
&\equiv& \Gamma(k)\,\frac{3}{5}\,\left\langle (I,J)_n\right|\sum_{i=1,3}\bfvec{\sigma}_{i}\cdot{\hat{\bf k}} \,\vec{\tau}_i\left|(1/2,1/2)\right\rangle\nonumber\\
\Gamma(k)&=& P(k)\,\frac{5}{3}\int d^3r\,\left[j_0(kr)\left(u^2(r)-\frac{v^2(r)}{3}\right)-\frac{4}{3}\,j_2(kr) \,v^2(r)\right].\label{FFSTATIC}
\end{eqnarray}
Notice that the form factor $\Gamma(k)$ is normalized such that $\Gamma(0)=g_A$.\\

The explicit form of the various contributions from the pionic sector are finally given by:
\begin{eqnarray}
E_{\pi M}&=&\frac{3}{2}\left(\frac{1}{2 F_\pi}\right)^2 \int{d^3k\over 
(2\pi)^3} \frac{{\bf k}^2 \,M^2_\pi}{2 \omega^2_k}\,\Gamma^2(k)\,\bigg[\frac{1}{\omega_k}\frac{1}{\omega_k+\epsilon_{N {\bf k}}}
+\frac{1}{(\omega_k+\epsilon_{N {\bf k}})^2}\nonumber\\
&&+\frac{32}{25}\,\bigg(\frac{1}{\omega_k}\frac{1}{\omega_k+\epsilon_{\Delta{\bf
k}}} + \frac{1}{(\omega_k+\epsilon_{\Delta{\bf k}})^2}\bigg)\bigg]\label{EPIMF}\\
E_{\pi K}&=&{3\over 2}\left({1\over 2 F_\pi}\right)^2 \int{d^3k\over 
(2\pi)^3} \frac{{\bf k}^4}{2 \omega^2_k}\,\Gamma^2(k)\,\bigg[\frac{1}{\omega_k}\frac{1}{\omega_k+\epsilon_{N {\bf k}}}
+\frac{1}{(\omega_k+\epsilon_{N {\bf k}})^2}\nonumber\\
&&+\frac{32}{25}\,\bigg(\frac{1}{\omega_k}\frac{1}{\omega_k+\epsilon_{\Delta{\bf
k}}} + \frac{1}{(\omega_k+\epsilon_{\Delta{\bf k}})^2}\bigg)\bigg]\label{EPIKF}\\
E_{\pi c}&=&-{3\over 2}\left({1\over 2 F_\pi}\right)^2 \int{d^3k\over 
(2\pi)^3} \frac{{\bf k}^2}{2}\,\Gamma^2(k)\,\bigg[\frac{1}{\omega_k}\frac{1}{\omega_k+\epsilon_{N {\bf k}}}
+\frac{1}{(\omega_k+\epsilon_{N {\bf k}})^2}\nonumber\\
&&+\frac{32}{25}\,\bigg(\frac{1}{\omega_k}\frac{1}{\omega_k+\epsilon_{\Delta{\bf
k}}} + \frac{1}{(\omega_k+\epsilon_{\Delta{\bf k}})^2}\bigg)\bigg]\nonumber\\
&&+\frac{3}{2}\left({1\over 2 F_\pi}\right)^2 \int{d^3k\over 
(2\pi)^3}\,k^2\,\Gamma^2(k)\,\bigg[\frac{1}{\omega_k}\frac{1}{\omega_k+\epsilon_{N {\bf k}}} +\frac{32}{25}\,\frac{1}{\omega_k}\frac{1}{\omega_k+\epsilon_{\Delta{\bf
k}}} \bigg]\label{EPICF}\\
E_{\pi Q}&=&-3\left({1\over 2 F_\pi}\right)^2 \int{d^3k\over 
(2\pi)^3} \,k^2\,\Gamma^2(k)\,\bigg[\frac{1}{\omega_k}\frac{1}{\omega_k+\epsilon_{N {\bf k}}} +\frac{32}{25}\,\frac{1}{\omega_k}\frac{1}{\omega_k+\epsilon_{\Delta{\bf
k}}} \bigg].\label{EPIQF}
\end{eqnarray}
Notice that, in the above expressions,  we have taken into account the recoil corrections with $\epsilon_{N \bfvec{k}}=\bfvec{k}^2/2 M_N$ and 
$\epsilon_{\Delta \bfvec{k}}=\omega_\Delta  +\bfvec{k}^2/2 M_\Delta$, $\omega_\Delta=M_\Delta -M_N$. A few remarks are in order. First, $E_{\pi M}$ is nothing but the pionic piece of the nucleon sigma term, which is expected to be  $\sigma_N^{(\pi)}\simeq \,20\,\rm{MeV}$. This light-quark sigma term has been discussed extensively in our previous papers \cite{Chanfray2007,Chanfray2011,Universe2025}. Second, we observe that in the static approximation - that is, when all recoil corrections are neglected  (i.e., $E_n\equiv 0$)- the energy $E_{\pi c}$, which is associated with the momentum conjugate of the pion field (i.e., its time derivative),  identically vanishes. Finally, the total contribution to the energy of pionic origin is:
\begin{eqnarray}
E_\pi &=& E_{\pi M}\,+\,E_{\pi K}\,+\,E_{\pi c}\,+\,E_{\pi Q}=  \frac{1}{2}  E_{\pi Q},\nonumber\\
&=& -\frac{3}{2}\,\left({1\over 2 F_\pi}\right)^2 \int{d^3k\over 
(2\pi)^3} \,k^2\,\Gamma^2(k)\,\bigg[\frac{1}{\omega_k}\frac{1}{\omega_k+\epsilon_{N {\bf k}}} +\frac{32}{25}\,\frac{1}{\omega_k}\frac{1}{\omega_k+\epsilon_{\Delta{\bf
k}}} \bigg].
\end{eqnarray}
\subsection{nucleon mass}
Before possible center of mass correction, the total energy of the nucleon is:
\begin{equation}
 E_0= E_Q \,+\, E_\pi=E_{QK}\,+E_{QM}\,+\, E_C\,+\,\frac{1}{2}  E_{\pi Q}.
\end{equation}
\\
Although the nucleon state, associated with a factorized wave function as a product of three quark orbitals (Eq.\eqref{LOCWF}), is not a physical state with fixed three-momentum, we will nonetheless attempt to model the (physical) nucleon using a localized state $|N: \bfvec{0}_X\rangle$ with inclusion of the pion cloud. In that respect, a rough estimate of the nucleon mass can be obtained by subtracting the spurious CM motion according to:
\begin{equation}
M_N=\sqrt{E_0^2 -\langle P^2\rangle} = \sqrt{E_0^2 -\frac{9}{2 b^2}}.
\end{equation}
For this first numerical estimate, we use the  NJLSet1 set of parameters \cite{Universe2025}, namely $M_q=356.7\,\rm{MeV}$, $M_\pi=140\,\rm{MeV}$, $F_\pi=91.7\,\rm{MeV}$ and the confining potential inspired from the  FCM framework (Eq. \eqref{FCMPOT}) with $\sigma=0.18\,\rm{GeV}^2$ and $T_g=0.283\,\rm{fm}$ \cite{Universe2025}.
The first conclusion that can be drawn from this study is that the finite size effect of the pion is absolutely required to ensure the mechanical stability of the nucleon, i.e., to satisfy the stability condition: 
\begin{eqnarray}
&&\frac{1}{3}\, E_{QK}\,-\, \frac{1}{3}\,P_C\,+\,\frac{1}{3}\, E_{\pi Q} \,+\, 
E_{\pi c} \,-\,\frac{1}{3}\,E_{\pi K} \,-\,E_{\pi M} =0.
\label{VONLAUEBIS}\end{eqnarray}
The value of the size parameter $c_\pi$, characterizing the extension of the non-local coupling of quarks to the pion field, must be determined as the outcome of a compromise. It must be large enough to prevent the quark core bag from collapsing. However, $c_\pi$
 should not be too large, to both ensure a sufficiently strong pionic attraction, resulting in a nucleon mass that is not too large, and to maintain a reasonably small quark core rms, meaning in practice not exceeding 0.5 fm. Taking  $c_\pi=0.8$ (i.e., the pion size parameter is $\rho_\pi=0.30\,\rm{fm}$),  we find that the von Laue condition is satisfied for a size parameter  $b_{virial}=0.345\,\rm{fm}$, leading to $r_B=0.495\,\rm{fm}$. The total energy before CM correction is $E_0=1.616\,\rm{GeV}$ leading to an estimated nucleon mass is  $E_0=\sqrt{E_0^2 -\langle P^2\rangle}=1.069\,\rm{GeV}$.  This value is reasonably close to the physical nucleon mass, especially considering that the perturbative gluon-exchange contribution, not considered here,  is expected to lower the mass by about  $- 100\, \rm{MeV}$ \cite{Universe2025}. However, an important drawback of these results is the too small value of the axial coupling constant, $g_A=1.13$, which is linked to the relatively small value of the size parameter  $b=0.345\,\rm{fm}$. This is most likely due to the large negative pressure from the confining force and the pion cloud, which must be counterbalanced by a large Fermi pressure, leading to a small value of the size parameter $b$. We will see, however, in the next section, how this can be improved by considering a physical state, $|N: \bfvec{0}_P\rangle$ for the nucleon with a well-defined momentum.
\subsection{Energy density and pressure distribution in  the static CCM nucleonic bag}
\subsubsection{Quark core contribution to the energy density and pressure  of the static CCM nucleonic bag }
The quark operators $\hat{T}_\Gamma$ entering   $\hat{T}^{ii}$ and $\hat{T}^{00}$ are $\mathcal{H}_{QK}$, $\mathcal{H}_{QM}$, $\mathcal{H}_{C}$ and $\mathcal{P}_{C}$. They are of the form $\hat{T}_\Gamma\sim \bar{q}(\bfvec{r})\Gamma (\bfvec{r})q(\bfvec{r})$. 
The ``static nucleonic bag" wave function is not translationally invariant and consequently the state $|N: \bfvec{0}_X\rangle=|N(\bfvec{X}_N=0)\rangle$ is not an eigenstate of the full momentum operator even if the average momentum is zero. In order to ensure  translational invariance in the evaluation of the static EMT, it is necessary to apply a projection method  \cite{Diakonov} that builds states of definite momentum $\bfvec{P}$  from the complete family of translated static bag states $|N(\bfvec{X}_N)\rangle$:
\begin{eqnarray}
t_\Gamma(\bfvec{r})&=& \int\frac{d^3\Delta\,d^3 y}{(2\pi)^3}  \,e^{-i\,(\bfvec{\Delta}\cdot(\bfvec{r}-\bfvec{y})} \,\left\langle\frac{\bfvec{\Delta}_N}{2}\left| \,\hat{T}_\Gamma(\bfvec{y})\,\right|-\frac{\bfvec{\Delta}_N}{2}\right\rangle\\
&=&\int\frac{d^3\Delta\,d^3 y}{(2\pi)^3}\,\int\frac{d^3 X_N}{V}  \,e^{-i\,(\bfvec{\Delta}\cdot(\bfvec{r}-\bfvec{y})} \,e^{-i\,\bfvec{\Delta}\cdot\bfvec{X}_N} \,\left\langle\bfvec{X}_N\left| \,\hat{T}_\Gamma(\bfvec{y})\,\right|\bfvec{X}_N\right\rangle\\
&=&\int\frac{d^3 X_N}{V}\left\langle\bfvec{X}_N\left| \,\hat{T}_\Gamma(\bfvec{r} +\bfvec{X_N})\,\right|\bfvec{X}_N\right\rangle\equiv\left\langle\bfvec{0}_X\left| \,\hat{T}_\Gamma(\bfvec{r} )\,\right|\bfvec{0}_X\right\rangle.
\end{eqnarray}
We  thus see that the result obtained using this projection procedure leads to the intuitive outcome anticipated at the beginning of Section~\ref{LOCAL}, where the spatial densities are radial functions ($r$ representing the distance from the center of the `static bag') given by the matrix elements of the various local operators (Eqs. \eqref{EPSQK}, \eqref{EPSQM}, \eqref{EPSQC}, and \eqref{EPSQCC})  between the QPW single-particle quark orbitals $\left(\psi\right)(\bfvec{r})$ explicited in \eqref{QPW2},\eqref{QPW1},\eqref{QPW3}, and \eqref{QPW4}:
\begin{eqnarray}
 \varepsilon_{QK}(r)/3&=&\left(\psi\right)^\dagger(\bfvec{r})[-i\bfvec{\alpha}\cdot\vec\nabla_{\bfvec{r}} ]\,\left(\psi\right)(\bfvec{r}) 
 = u(r)v'(r)-u'(r) v(r)\,+\,\frac{2 u(r) v(r)}{r}\,\\
 \varepsilon_{QM}(r)/3&=&\left(\psi\right)^\dagger(\bfvec{r})\, \beta \,M_q\,\left(\psi\right)(\bfvec{r}) =M_q\,\left(u^2(r) -v^2(r)\right)\\
 \varepsilon_{C}(r)/3&=&\left(\psi\right)^\dagger(\bfvec{r})\,W_C(\bfvec{r})\,\left(\psi\right)(\bfvec{r}) = W_C(r)\,\left(u^2(r) +v^2(r)\right)\\
 -p_{C}(r)/3&=&-\left(\psi\right)^\dagger(\bfvec{r})\, \bfvec{}\cdot\bfvec{\nabla}W_C(\bfvec{r})\,\left(\psi\right)(\bfvec{r}) =-r\, W'_C(r)\,\left(u^2(r) +v^2(r)\right).
\end{eqnarray}
Hence the quark core contributions to the energy density and pressure are: 
\begin{eqnarray}
\varepsilon_{Q}(r)=\varepsilon_{QK}(r)\,+\,\varepsilon_{QM}(r)\,+\,\varepsilon_C(r)\\
p_Q(r)=\frac{1}{3}\varepsilon_{QK}(r)\,-\,\frac{1}{3} p_C(r).
\end{eqnarray}
\subsubsection{Pion cloud contribution to the energy density and pressure of the ``static CCM nucleonic bag" }
The calculation of the pion cloud contribution to the static EMT tensor also necessitates a projection procedure. To this end, let us introduce the following Green's function:
\begin{equation}
G_a(\bfvec{r},t\,;\, \bfvec{r}',t' )=\sum_{\bfvec{\Delta}}\, \left\langle\bfvec{\Delta}_N/2\left|\,-i \,\mathcal{T}\left(\Phi_a(\bfvec{r},t)\,,\,\Phi_a(\bfvec{r}',t')\,\right)\right|-\bfvec{\Delta}_N/2\right\rangle.
\end{equation}
From the equation of motion for the pion field given in Eq.~\eqref{EOMPI} we can immediately deduce: 
\begin{eqnarray}
&&\left( -\frac{\partial^2 }{\partial t^2} +\nabla_{\bfvec{r}}^2 - M^2_\pi\right) G_a(\bfvec{r},t\,;\, \bfvec{r'},t' )\left( -\frac{\partial^2 }{\partial t'^2} +\nabla^2_{\bfvec{r}'} - M^2_\pi\right)
\nonumber\\
&&\quad=\left(\frac{1}{2 F_\pi}\right)^2\sum_{\bfvec{\Delta}}\, \left\langle\bfvec{\Delta}_N/2\left|\,-i \,\mathcal{T}\left(\bfvec{\nabla}_{\bfvec{r}}\cdot\tilde{\bfvec{S}}^\dagger_a(\bfvec{r},t)\,,\,\bfvec{\nabla}_{\bfvec{r}'}\cdot\tilde{\bfvec{S}}_a (\bfvec{r}',t')\right)\right|-\bfvec{\Delta}_N/2\right\rangle .
\end{eqnarray}
Hence: 
\begin{eqnarray}
 &&G_a(\bfvec{r},t\,;\,\bfvec{r}',t')=\left(\frac{1}{2 F_\pi}\right)^2\int dt_1\,d^3r_1 \,dt'_1\,d^3r'_1,\,\int\frac{d\omega}{2\pi} \int\frac{d^3 k}{(2\pi)^3}\,e^{-i\omega(t-t_1)}\,e^{i\bfvec{k}\cdot(\bfvec{r}-\bfvec{r}_1)} \nonumber\\
 &&\int\frac{d\omega'}{2\pi} \int\frac{d^3 k'}{(2\pi)^3}\,e^{i\omega'(t'-t'_1)} e^{-i\bfvec{k}'\cdot(\bfvec{r}'-\bfvec{r}'_1)}\frac{\sum_{\bfvec{\Delta}} \langle\bfvec{\Delta}_N/2|-i \,\mathcal{T}\left(\bfvec{\nabla}_{\bfvec{r}_1}\cdot\tilde{\bfvec{S}}^\dagger_a(\bfvec{r}_1,t_1)\,,\,\bfvec{\nabla}_{\bfvec{r}_1'}\cdot\tilde{\bfvec{S}}_a (\bfvec{r}_1',t_1')\right)|-\bfvec{\Delta}_N/2\rangle}{(\omega^2-\omega_k^2 +i\eta)\,({\omega'}^{2}-{\omega'}_k^2 +i\eta)}.\label{GaFIRST}
\end{eqnarray}
We now introduce a polarization propagator:
\begin{eqnarray}
 \Pi_{0D}(\bfvec{k},\bfvec{q};t_1,t_1')&=&  \sum_{\bfvec{\Delta}}\, \langle\bfvec{\Delta}_N/2|\,-i \,\mathcal{T}\left(\tilde{S}^\dagger_{\bfvec{k}}(t_1)\,,\,\tilde{S}_{\bfvec{q}}(t'_1)\right)|-\bfvec{\Delta}_N/2\rangle =\int\frac{d\nu}{2\pi}\,e^{-i\nu(t_1-t'_1)}\,\Pi_{0D}(\bfvec{k},\bfvec{q};\nu)\nonumber\\
 \Pi_{0D}(\bfvec{k},\bfvec{q};\nu)&=&\sum_{n,\bfvec{\Delta}}\left(\frac{ \langle\bfvec{\Delta}_N/2|\,\tilde{S}^\dagger_{\bfvec{k}}\,|n\rangle\,\langle n|\,\tilde{S}_{\bfvec{q}}\,|-\bfvec{\Delta}_N/2\rangle }{\nu-E_n+i\eta}\,-\,
\frac{\langle\bfvec{\Delta}_N/2|\,\tilde{S}_{\bfvec{q}}\,|n\rangle\,\langle n|\,\tilde{S}^\dagger_{\bfvec{k}}\,|-\bfvec{\Delta}_N/2\rangle}{\nu +E_n -i\eta}\right).
\end{eqnarray}
To avoid possible confusion between the numerous $\bfvec{k}$ and $\bfvec{k}'$, we replace in Eq. \eqref{GaFIRST} the $\bfvec{k}'$ integration variable with $\bfvec{q}$. After performing the time integration, we arrive  at:
\begin{eqnarray}
G_a(\bfvec{r},t\,;\,\bfvec{r},t')&=& \left(\frac{1}{2 F_\pi}\right)^2 \int\frac{d^3 k}{(2\pi)^3} \frac{d^3 q}{(2\pi)^3}\,e^{i\,(\bfvec{k}-\bfvec{q})\cdot\bfvec{r}}\, \int\frac{d\omega}{2\pi}\,e^{-i\omega(t-t_1)}\,\frac{\Pi_{0D}(\bfvec{k},\bfvec{q};\omega)}{(\omega^2-\omega_k^2 +i\eta)\,(\omega^2-\omega_q^2 +i\eta)}.
\end{eqnarray}
The spatial distribution  associated with the pionic mass density operator can now be written as follows: 
\begin{eqnarray}
 \varepsilon_{\pi M}(r)&=&\frac{M^2_\pi}{2} \, \sum_{\bfvec{\Delta}}\langle\bfvec{\Delta}_N/2|\, \mathcal{H}_{\pi M}(0,\bfvec{r})\,|-\bfvec{\Delta}_N/2\rangle\nonumber\\
 &=&\frac{M^2_\pi}{2} \lim_{t'\to t^+}\sum_{a,\bfvec{\Delta}}\langle\bfvec{\Delta}_N/2|\mathcal{T}\left( \Phi_a(\bfvec{r},t)\,\Phi_a(\bfvec{r},t')\right)|-\bfvec{\Delta}_N/2\rangle=\frac{i\,M^2_\pi}{2}\sum_{a}\,\lim_{t'\to t^+}G_a(\bfvec{r},t\,;\,\bfvec{r},t')\nonumber\\
&=& \frac{3}{2}\left(\frac{M_\pi}{2 F_\pi}\right)^2 \int\frac{d^3 k}{(2\pi)^3} \frac{d^3 q}{(2\pi)^3}\,e^{i\,(\bfvec{k}-\bfvec{q})\cdot\bfvec{r}}\, \int\frac{d\omega}{2\pi}\,e^{i\omega\eta^+}\,\frac{\Pi_{0D}(\bfvec{k},\bfvec{q};\omega)}{(\omega^2-\omega_k^2 +i\eta)\,(\omega^2-\omega_q^2 +i\eta)}.\label{DENSPIM0}
\end{eqnarray}
For given values of $\bfvec{k}$ and $\bfvec{q}$, $\Pi_{0D}(\bfvec{k},\bfvec{q};\omega)$  involves a double summation over the momentum  $\bfvec{\Delta}$ and over the intermediate states $|n\rangle$ (in practice nucleon and $\Delta(1232)$ states), of the type: 
$$\sum_{n,\bfvec{\Delta}}F(E_n)\,\langle\bfvec{\Delta}_N/2|\,\tilde{S}^\dagger_{\bfvec{k}}\,|n\rangle\,\langle n|\,\tilde{S}_{\bfvec{q}}\,|-\bfvec{\Delta}_N/2\rangle .$$
Momentum conservation implies $\bfvec{P}_n=\bfvec{q}-\bfvec{\Delta}/2=\bfvec{k}+\bfvec{\Delta}/2$. Thus, for given $\bfvec{k}$ and $\bfvec{q}$, both $\bfvec{\Delta}$ and the momentum and energy  of the intermediate states are fixed:
$$\bfvec{\Delta}=\bfvec{q}-\bfvec{k},\qquad \bfvec{P}_n=\frac{\bfvec{q}+\bfvec{k}}{2},\qquad
E_{nN}=\frac{(\bfvec{q}+\bfvec{k})^2}{4M_N},\qquad
E_{n\Delta}=\omega_\Delta +\frac{(\bfvec{q}+\bfvec{k})^2}{4M_\Delta}.$$
The simplest way to account for translational invariance is to adopt the following prescription: 
\begin{eqnarray}
&&\langle (I,J)_n: \bfvec{P}_n =\bfvec{q}-\bfvec{\Delta}/2\,|\,\tilde{S}_{\bfvec{q}}\,|-\bfvec{\Delta}/2\rangle = 
\left\langle \bfvec{0}_X :(I,J)_n \right|\,\tilde{S}_{\bf q}\, \left|\bfvec{0}_X\right\rangle\nonumber\\
&& =\Gamma(q)\,\frac{3}{5}\,\left\langle (I,J)_n\right|\sum_{i=1,3}\bfvec{\sigma}_{i}\cdot{\hat{\bf q}} \,\tau_{3 i}\left|(1/2,1/2)\right\rangle\equiv\Gamma(q)\,\langle \bfvec{\sigma}\cdot{\hat{\bf q}}\rangle_n ,
\end{eqnarray}
where $\Gamma(q)$ is the static form factor, whose explicit expression is given in Eq. \eqref{FFSTATIC}. As in Section \ref{PIENERGY}, the energy integration in Eq. \eqref{DENSPIM0} can be performed with a Wick rotation. The result is: 
\begin{eqnarray}
 \varepsilon_{\pi M}(r)&=&\frac{3}{2} \left(\frac{1}{2 F_\pi}\right)^2 \int\frac{d^3 k}{(2\pi)^3} \frac{d^3 q}{(2\pi)^3}\,e^{i\,(\bfvec{k}-\bfvec{q})\cdot\bfvec{r}}\,\Gamma(k) \,\Gamma(q)\,\frac{M^2_\pi\,k\,q}{\omega_k\omega_q\,(\omega_k+\omega_q)}\nonumber\\
 &&\qquad\sum_n\frac{E_n +\omega_k+\omega_q}{(E_n+\omega_k)(E_n+\omega_q)}\,\langle \bfvec{\sigma}\cdot{\hat{\bf k}}\rangle_n\langle \bfvec{\sigma}\cdot{\hat{\bf q}}\rangle_n.
\end{eqnarray}
Notice that the volume integration of Eq. \eqref{DENSPIM0} allows one to recover the pion mass energy, $E_{\pi M}=\int d^3 r\,\varepsilon_{\pi M}(r)$ as written in Eq.\eqref{EPIM}, and automatically selects $\bfvec{k}=\bfvec{q}$. For the practical numerical calculation, we replace $E_{n\bfvec{k}\bfvec{q}}=\epsilon_{N,\Delta}((\bfvec{k}+\bfvec{q})/2)=\bar{\epsilon}_{N,\Delta \bfvec{k}\bfvec{q}}$ with $\bar{\epsilon}_{N,\Delta kq}=(E_n(k)+E_n(q))/2$, in order  to avoid an additional angular integration, without affecting the total energy. Performing the various remaining angular integrations, one obtains an explicit form for the pion mass energy density:
\begin{eqnarray}
 \varepsilon_{\pi M}(r)&=&\frac{3}{2} \left(\frac{1}{2 F_\pi}\right)^2 \int\frac{d^3 k}{(2\pi)^3} \frac{d^3 q}{(2\pi)^3}\,j_1(kr)\,j_1(qr)\,\Gamma(k) \,\Gamma(q)\, \frac{M^2_\pi\,k\,q}{2\omega_k\omega_q}\nonumber\\
&& \bigg[\frac{1}{(\omega_k+\omega_q)(\omega_k+\bar{\epsilon}_{Nkq})}+\frac{1}{(\omega_k+\omega_q)(\omega_q+\bar{\epsilon}_{Nkq})}+\frac{1}{(\omega_k+\bar{\epsilon}_{Nkq})(\omega_q+\bar{\epsilon}_{Nkq})}
\nonumber\\
&&+\frac{32}{25}\,\bigg(\frac{1}{(\omega_k+\omega_q)(\omega_k+\bar{\epsilon}_{\Delta kq})}+\frac{1}{(\omega_k+\omega_q)(\omega_q+\bar{\epsilon}_{\Delta kq})}+\frac{1}{(\omega_k+\bar{\epsilon}_{\Delta kq})(\omega_q+\bar{\epsilon}_{\Delta kq})}\bigg)\bigg].\label{DPIM}
\end{eqnarray}
The pion kinetic energy distribution can be obtained in a similar manner:
\begin{eqnarray}
 \varepsilon_{\pi K}(r)&=&\frac{3}{2} \left(\frac{1}{2 F_\pi}\right)^2 \int\frac{d^3 k}{(2\pi)^3} \frac{d^3 q}{(2\pi)^3}\,e^{i\,(\bfvec{k}-\bfvec{q})\cdot\bfvec{r}}\,\Gamma(k) \,\Gamma(q)\,\frac{k^2\,q^2}{\omega_k\omega_q\,(\omega_k+\omega_q)}\nonumber\\
 &&\qquad\sum_n\frac{E_n +\omega_k+\omega_q}{(E_n+\omega_k)(E_n+\omega_q)}\,\,{\hat{\bf k}}\cdot{\hat{\bf q}}\,\langle \bfvec{\sigma}\cdot{\hat{\bf k}}\rangle_n\langle \bfvec{\sigma}\cdot{\hat{\bf q}}\rangle_n,
\end{eqnarray}
with explicit form: 
\begin{eqnarray}
 \varepsilon_{\pi K}(r)&=&\frac{3}{2} \left(\frac{1}{2 F_\pi}\right)^2 \int\frac{d^3 k}{(2\pi)^3} \frac{d^3 q}{(2\pi)^3}\,\left(\frac{1}{3}j_0(kr)\,j_0(qr)\,+\,\frac{2}{3}j_2(kr)\,j_2(qr)\right)\,\Gamma(k) \,\Gamma(q)\, \frac{k^2\,q^2}{2\omega_k\omega_q}\nonumber\\
&& \bigg[\frac{1}{(\omega_k+\omega_q)(\omega_k+\bar{\epsilon}_{Nkq})}+\frac{1}{(\omega_k+\omega_q)(\omega_q+\bar{\epsilon}_{Nkq})}+\frac{1}{(\omega_k+\bar{\epsilon}_{Nkq})(\omega_q+\bar{\epsilon}_{Nkq})}
\nonumber\\
&&+\frac{32}{25}\,\bigg(\frac{1}{(\omega_k+\omega_q)(\omega_k+\bar{\epsilon}_{\Delta kq})}+\frac{1}{(\omega_k+\omega_q)(\omega_q+\bar{\epsilon}_{\Delta kq})}+\frac{1}{(\omega_k+\bar{\epsilon}_{\Delta kq})(\omega_q+\bar{\epsilon}_{\Delta kq})}\bigg)\bigg].\label{DPIK}
\end{eqnarray}
To obtain the spatial energy density associated with the pion conjugate momentum, one simply has to replace in Eq. \eqref{DENSPIM0} $M^2_\pi$ by an extra energy factor $\omega^2$. By following the same steps as for $\varepsilon_{\pi M}$, we obtain:
\begin{eqnarray}
 \varepsilon_{\pi c}(r)&=&-\frac{3}{2} \left(\frac{1}{2 F_\pi}\right)^2 \int\frac{d^3 k}{(2\pi)^3} \frac{d^3 q}{(2\pi)^3}\,j_1(kr)\,j_1(qr)\,\Gamma(k) \,\Gamma(q)\, \frac{k\,q\,(\omega^2_k+\omega^2_q)}{4\omega_k\omega_q}\nonumber\\
&& \bigg[\frac{1}{(\omega_k+\omega_q)(\omega_k+\bar{\epsilon}_{Nkq})}+\frac{1}{(\omega_k+\omega_q)(\omega_q+\bar{\epsilon}_{Nkq})}+\frac{1}{(\omega_k+\bar{\epsilon}_{Nkq})(\omega_q+\bar{\epsilon}_{Nkq})}
\nonumber\\
&&+\frac{32}{25}\,\bigg(\frac{1}{(\omega_k+\omega_q)(\omega_k+\bar{\epsilon}_{\Delta kq})}+\frac{1}{(\omega_k+\omega_q)(\omega_q+\bar{\epsilon}_{\Delta kq})}+\frac{1}{(\omega_k+\bar{\epsilon}_{\Delta kq})(\omega_q+\bar{\epsilon}_{\Delta kq})}\bigg)\bigg]\nonumber\\
&&+\frac{3}{2} \left(\frac{1}{2 F_\pi}\right)^2 \int\frac{d^3 k}{(2\pi)^3} \frac{d^3 q}{(2\pi)^3}\,j_1(kr)\,j_1(qr)\,\Gamma(k) \,\Gamma(q)\,\frac{k\,q}{2} \nonumber\\&&\bigg[\frac{1}{\omega_k}\frac{1}{\omega_k+\bar{\epsilon}_{Nkq}} + \frac{1}{\omega_q}\frac{1}{\omega_q+\bar{\epsilon}_{Nkq}} _,+\,\frac{32}{25}\,\left(\frac{1}{\omega_k}\frac{1}{\omega_k+\bar{\epsilon}_{\Delta kq}}+\frac{1}{\omega_q}\frac{1}{\omega_q+\bar{\epsilon}_{\Delta kq}}\right) \bigg].\label{DPIC}
\end{eqnarray}
A similar Green's function approach allows us to calculate the spatial density associated with the pion-quark density operator $\mathcal{H}_{\pi Q}$,
\begin{eqnarray}
 \varepsilon_{\pi Q}(r)&=&-3\left(\frac{1}{2 F_\pi}\right)^2\int\frac{d^3 k}{(2\pi)^3}   \,k^2\,\,\left(j_0(kr)\,\tilde{\Gamma}_0(r)\,+\,j_2(kr)\,\tilde{\Gamma}_2(r)\right)\,\Gamma(k)\nonumber\\
 &&\bigg[\frac{1}{\omega_k}\frac{1}{\omega_k+\epsilon_{N {\bf k}}} +\frac{32}{25}\,\frac{1}{\omega_k}\frac{1}{\omega_k+\epsilon_{\Delta{\bf
k}}} \bigg],\label{DPIQ}   
\end{eqnarray}
with:
\begin{eqnarray}
 \tilde{\Gamma}_0(r)&=&\int \frac{d^3 t}{(2\pi)^3}\, j_0(tr)\,\Gamma_0(t)\\
 \tilde{\Gamma}_2(r)&=&\int \frac{d^3 t}{(2\pi)^3}\, j_2(tr)\,\Gamma_2(t)\\
 \Gamma_0(t)&=& P(t)\,\frac{5}{3}\int d^3x\,\,j_0(tx) \,\left(u^2(x)-\frac{v^2(x)}{3}\right)\\
 \Gamma_2(t)&=& P(t)\,\frac{5}{3}\left(-\frac{4}{3}\int d^3x\,\,j_2(tx) \,v^2(x)\right)\\
 \Gamma(k)&=& P(k)\,\frac{5}{3}\int d^3x\,\left[j_0(kx)\left(u^2(x)-\frac{v^2(x)}{3}\right)-\frac{4}{3}\,j_2(kx) \,v^2(x)\right].
\end{eqnarray}
\begin{figure}
\includegraphics[width=0.45\textwidth,angle=0]{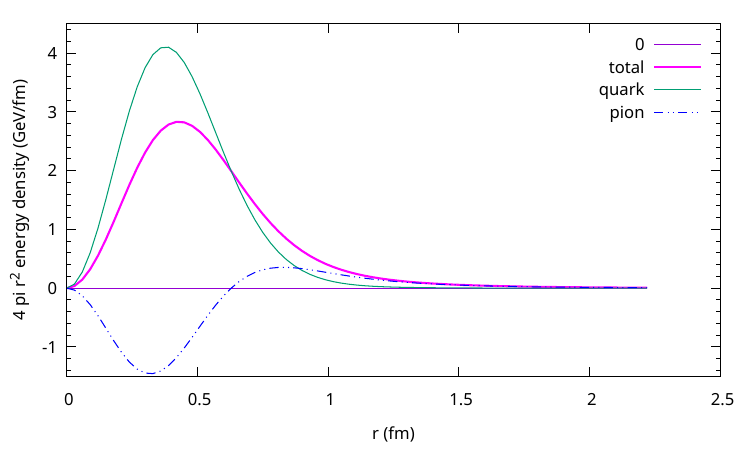}|
\includegraphics[width=0.45\textwidth,angle=0,]{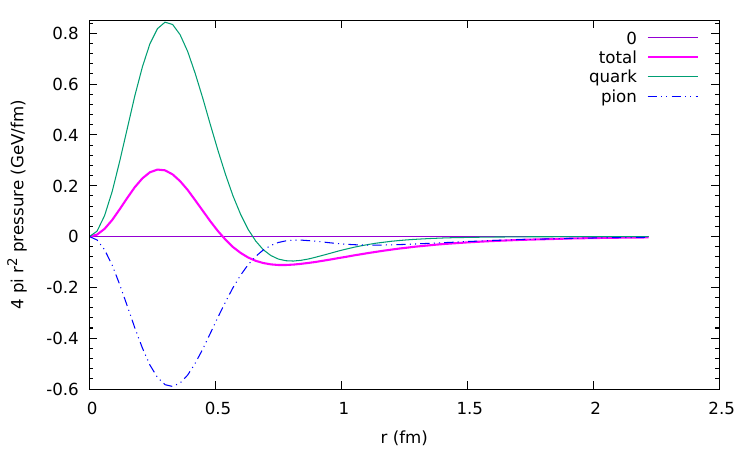}
\caption{Left panel: energy density distribution, $\varepsilon(r)=\varepsilon_{QK}(r)\,+\,\varepsilon_{QM}(r)\,+ \varepsilon_C(r)\,+\, \varepsilon_{\pi Q}(r) \,+\, 
\varepsilon_{\pi c}(r) \,+\,\varepsilon_{\pi K}(r) \,+\varepsilon_{\pi M}(r) $, as a function of $r$ multiplied by $4\pi r^2$ for the localized static nucleonic bag. Also shown are the separate quark and pionic distributions. Right panel: pressure distribution, $p(r)=\frac{1}{3}\varepsilon_{QK}(r)\,-\frac{1}{3} p_C(r)\,+\,\frac{1}{3} \varepsilon_{\pi Q}(r) \,+\, 
\varepsilon_{\pi c}(r) \,-\,\frac{1}{3}\varepsilon_{\pi K}(r) \,-\varepsilon_{\pi M}(r)$,  as a function of $r$ multiplied by $4\pi r^2$. Also shown are the separate quark (i.e., non pionic) and pionic distributions.}
\label{NOCMENERGY}
\end{figure}
 Using the same input parameters  as for the calculation of the total energy, the resulting radial energy density and pressure  distribution are shown in Fig. \ref{NOCMENERGY}.
\section{Momentum projected nucleon trial wave function and center of mass correction}
As  announced before, we now replace the factorized wave function, associated with the state, $|N: \bfvec{0}_X\rangle=|N(\bfvec{X}_N=0)\rangle$, with a translationally invariant nucleon wave function, associated with the state, $|N: \bfvec{0}_P\rangle=|N(\bfvec{P}_N=0)\rangle$, i.e., a state projected onto fixed three-momentum:
\begin{eqnarray}
|N(\bfvec{P}_N)\rangle &=& V^{3/2}\,\int \frac{d^3 p_1}{(2\pi)^3}\frac{d^3 p_2}{(2\pi)^3}\frac{d^3 p_3}{(2\pi)^3}\,\frac{N_{\bf{P}_N}}{V^{1/2}} \, \Phi\left({\bf p}_1\right)\,\Phi\left({\bf p}_2\right)\,  \Phi\left({\bf p}_3\right)\nonumber\\
&&\qquad\qquad(2\pi)^3\,\delta \left(\bf{P}_N - \bf{p}_1 -\bf{p}_2 - \bf{p}_3 \right)\,B^\dagger_{{\bf p}_1}
 B^\dagger_{{\bf p}_2}(\mathcal{S}) B^\dagger_{{\bf p}_3}(\mathcal{S})\,\,\left|\varphi(\mathcal{S})\right\rangle \nonumber\\
 &=&\frac{N_{\bf{P}_N}}{V^{1/2}}\,\int d^3X_N\,|N(\bfvec{X}_N)\rangle\,\,e^{i\bfvec{P}_N \cdot \bfvec{X}_N }.
\end{eqnarray}
Let us now outline how the calculations of the various quantities are modified.
\subsection{Form factors}
Let us consider again the matrix elements of a local quark bilinear operator (with $\Gamma$ being some matrix in Dirac-flavor space) between two states with well defined momentum, ie, $|N(\bfvec{P}_N)\rangle\equiv|\bfvec{P}_N\rangle$. From translational invariance, such a matrix element should take the form:
\begin{equation}
 \langle\bfvec{p}'| \bar{q}(\bfvec{r})\,\Gamma\,q(\bfvec{r})|\bfvec{p}\rangle=\frac{1}{V}\, e^{-i\,(\bfvec{p}' -\bfvec{p})\cdot\bfvec{r}}\,\Gamma_N \left(\bfvec{p}'-\bfvec{p}\right)   \,\bar{\eta}_N \,(\bfvec{p}')\Gamma \,\eta_N(\bfvec{p}),
\end{equation}
where again all the states are normalized to one in a large box of volume $V$. However, when the matrix element is calculated between the states $|N(\bfvec{P}_N=\bfvec{p})\rangle$ and $|N(\bfvec{P}_N=\bfvec{p}')\rangle$, the form factor depends separately on $\bfvec{p}'$ and $\bfvec{p}$ and not only on $\bfvec{\Delta}=\bfvec{p}'$\,-\,$\bfvec{p}$. In the context of calculating  the static EMT tensor, it is natural to compute these form factors in the Breit Frame where the local  bilinear quark operator transfers only three-momentum to the nucleon. Thus the prescription is: 
\begin{equation}
 \langle\bfvec{p}'|\, \bar{q}(\bfvec{r})\,\Gamma\,q(\bfvec{r})\,|\bfvec{p}\rangle=\langle \bfvec{\Delta}_N/2|\,\bar{q}(\bfvec{r})\,\Gamma\,q(\bfvec{r})|-\bfvec{\Delta}_N/2\,\rangle\qquad\hbox{with}\qquad \bfvec{\Delta}=\bfvec{p}' \,-\,\bfvec{p}.
\end{equation}
Using the (normalized) Gaussian single-quark Fock-Space amplitude $\Phi(\bfvec{p})\sim\exp{-b^2p^2/2}$ \eqref{TRIALWF}, one can analytically demonstrate the following result:
\begin{eqnarray}
\langle \bfvec{\Delta}_N/2|\,\bar{q}(\bfvec{r})\,\Gamma\,q(\bfvec{r})|-\bfvec{\Delta}_N/2\rangle&=&\frac{1}{V}\, e^{-i\,\bfvec{\Delta} \cdot\bfvec{r}}\, e^{5 b^2\Delta^2/24}\,\int d^3x\,e^{i\,\bfvec{\Delta} \cdot\bfvec{x}} \,\left(\bar{\psi}\right)_E({\bf{x}})\,\Gamma\,\left(\psi\right)_E({\bf{x}}),
\end{eqnarray}
where $\left(\bar{\psi}\right)_E({\bf{x}})$ is the QPW wave function defined in Eqs.  \eqref{QPW2},\eqref{QPW1},\eqref{QPW3},\eqref{QPW4}, \eqref{TRIALWF},\ but with the  parameter $b$ replaced with an effective parameter $b_E$ such that:
\begin{equation}
 \left(\psi\right)({\bf{x}}) \,\to\,   \left(\psi\right)_E({\bf{x}})\qquad\Leftrightarrow\qquad b^2\,\to\, b^2_E=\frac{3}{2}\,b^2.
\end{equation}
The replacement of the factorized wave function by the translationally invariant one leads to the following immediate consequences:
\begin{itemize}
\item 
The pionic form factors $\Gamma_\pi(k)$ appearing in the various expressions relative to the pionic energies, Eqs. \eqref{EPIMF}\,\eqref{EPIKF}, \eqref{EPICF}, \eqref{EPIQF}, and pionic energy densities, Eqs. \eqref{DPIM}\,\eqref{DPIK}, \eqref{DPIC}, \eqref{DPIQ}, are obtained by replacing $b$ by $b_E$ and multiplying them by $exp\,(5 b^2k^2/24)$. For example, the pionic form factor (Eq.~\eqref{FFSTATIC}) becomes:
\begin{equation}
 \Gamma(k)=  e^{5 b^2k^2/24}\,P(k)\,\frac{5}{3}\int d^3x\,\left[j_0(kx)\left(u_E^2(x)-\frac{v_E^2(x)}{3}\right)-\frac{4}{3}\,j_2(kx) \,v_E^2(x)\right].   
\end{equation}
This is all we need for the calculation of energies of pionic origin and the associated spatial densities.
\item 
As a corollary, the nucleon axial coupling constant becomes:
\begin{equation}
 g_A =  \frac{5}{3}\,\int d^3x\,\left(u_E^2(x)-\frac{v_E^2(x)}{3}\right)\label{GAEX}.
\end{equation}
\item
The baryon number radius is modified according to:
\begin{equation}
 \left\langle r^2_B\right\rangle =  \int d^3x\,x^2\,\left(u_E^2(x)+v_E^2(x)\right)\,
 \,-\,\frac{5}{4}\,b^2=\left\langle r^2_B\right\rangle_E \,-\,\frac{5}{6}\,b_E^2 .\label{RBEX2}
\end{equation}

    \end{itemize}
 
\subsection{Quark number density}
As already introduced and justified  in the introduction, the quark number density is defined in the Wigner sense  as: 
\begin{eqnarray}
\rho_B(r)&=&\int\frac{V d^3\Delta}{(2\pi)^3}\, \langle \bfvec{\Delta}_N/2|\,q^\dagger(\bfvec{r})\,q(\bfvec{r})|N(-\bfvec{\Delta}_N/2)\rangle\nonumber\\
&=&\int\frac{d^3\Delta}{(2\pi)^3}\, e^{-i\,\bfvec{\Delta} \cdot\bfvec{r}}\, e^{5 b^2\Delta^2/24}\,\int d^3x\,e^{i\,\bfvec{\Delta} \cdot\bfvec{x}} \,\left(\psi\right)^\dagger_E({\bf{x}})\,\left(\psi\right)_E({\bf{x}})\\
&=&\int\frac{d^3\Delta\,d^3x}{(2\pi)^3}\, e^{5 b^2\Delta^2/24}\,j_0(\Delta r)\,j_0(\Delta x)\,
\left(\psi^\dagger_E \psi_E\right)(x),\label{WIGN1}
\end{eqnarray}
with the normalization condition:
\begin{equation}
\int d^3 r  \rho_B(r) =1.
\end{equation}
\subsection{quark kinetic and mass energy and associated energy densities}
\begin{figure}
\includegraphics[width=0.8\textwidth,angle=0]{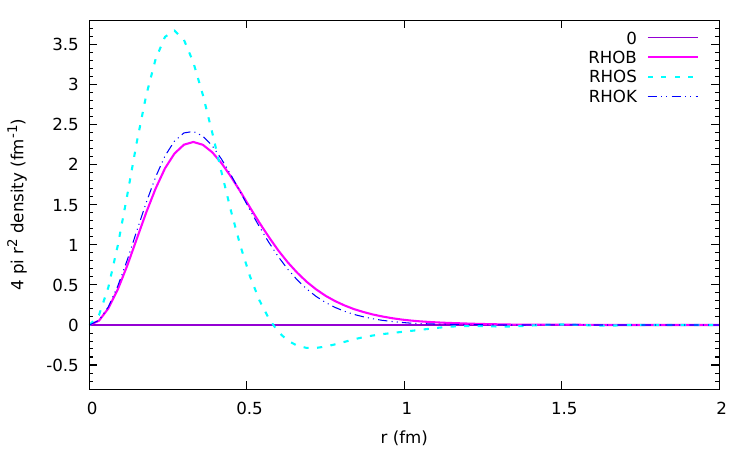}|
\caption{The normalized quark number density $\rho_B(r)$, quark mass energy $\rho_S(r)$ and quark kinetic energy $\rho_K(r)$ distributions.
The reader may notice a slight oscillatory behavior at large values of $r$, which is an artifact of the delicate double numerical integration of the product of two Bessel functions, which may become artificially divergent at large distances. 
In the present work we have chosen a large-distance cutoff such that the sum rule, $\int d^3r\,\rho_{B,S,K}(r)=1$, is satisfied to better than one part in a thousand for each of the three densities.}
\label{WIGNER}
\end{figure}

 The total quark kinetic energy and quark mass energy are obtained from their expression for the localized static CCM bag by simply replacing the parameter $b$ with $b_E$:
 \begin{eqnarray}
 E_{QK}&=& 3\int d^3 r \left(\psi\right)^\dagger_E({\bf{r}}) \,  (-i\bfvec{\alpha}\cdot\vec\nabla_{\bfvec{r}})\,\left(\psi\right)_E({\bf{r}}),\label{WIGN2}\\
 E_{QM}&=& 3\int d^3 r \left(\bar{\psi}\right)_E({\bf{r}}) \,(\beta \,M_q \,\left(\psi\right)_E({\bf{r}}).\label{WIGN3}
 \end{eqnarray}
 The associated spatial  distributions are given by:
 \begin{eqnarray}
\varepsilon_{QK}(r)&=&3\int\frac{V d^3\Delta}{(2\pi)^3}\, \langle \bfvec{\Delta}_N/2|\,q^\dagger(\bfvec{r})\,(-i\bfvec{\alpha}\cdot\vec\nabla_{\bfvec{r}})\,q(\bfvec{r})|-\bfvec{\Delta}_N/2\rangle\nonumber\\
&=&\int\frac{d^3\Delta}{(2\pi)^3}\, e^{-i\,\bfvec{\Delta} \cdot\bfvec{r}}\, e^{5 b^2\Delta^2/24}\,\int d^3x\,e^{i\,\bfvec{\Delta} \cdot\bfvec{x}} \,\left(\psi\right)^\dagger_E({\bf{x}})\,(-i\bfvec{\alpha}\cdot\vec\nabla_{\bfvec{x}})\left(\psi\right)_E({\bf{x}})\nonumber\\
&=&\int\frac{d^3\Delta\,d^3x}{(2\pi)^3}\, e^{5 b^2\Delta^2/24}\,j_0(\Delta r)\,j_0(\Delta x)\,
\left(\psi^\dagger_E \,(-i\bfvec{\alpha}\cdot\vec\nabla_{\bfvec{x}})\,\psi_E\right)(x)\nonumber\\
&\equiv& E_{QK}\,\rho_K(r) \label{EQKW}\\
\varepsilon_{QM}(r)&=&3\int\frac{V d^3\Delta}{(2\pi)^3}\, \langle \bfvec{\Delta}_N/2|\,q^\dagger(\bfvec{r})\,(\beta\,M_q)\,q(\bfvec{r})|-\bfvec{\Delta}_N/2\rangle\nonumber\\
&=&\int\frac{d^3\Delta}{(2\pi)^3}\, e^{-i\,\bfvec{\Delta} \cdot\bfvec{r}}\, e^{5 b^2\Delta^2/24}\,\int d^3x\,e^{i\,\bfvec{\Delta} \cdot\bfvec{x}} \,\left(\psi\right)^\dagger_E({\bf{x}})\,(\beta\,M_q)\left(\psi\right)_E({\bf{x}})
\nonumber\\
&=&\int\frac{d^3\Delta\,d^3x}{(2\pi)^3}\, e^{5 b^2\Delta^2/24}\,j_0(\Delta r)\,j_0(\Delta x)\,
\left(\psi^\dagger_E \,(\beta\,M_q)\,\psi_E\right)(x),\nonumber\\
&\equiv& E_{QM}\,\rho_S(r), 
\label{EQMW}
\end{eqnarray}
 and they satisfy:
 \begin{equation}
 \int d^3 r \, \varepsilon_{QK}(r) =E_{QK},\qquad\qquad  \int d^3 r \, \varepsilon_{QM}(r) =E_{QM}.
 \end{equation}
 In the above equations, we have introduced the normalized kinetic energy $\rho_K(r)$ and quark mass $\rho_S(r)$ Wigner densities. The three Wigner densities are plotted in Fig~\ref{WIGNER} for a value of the size parameter $b_E=0.423\,\rm{fm}$ corresponding to a quark core radius $r_B=0.45\, \rm{fm}$. 
\subsection{energy, energy density and pressure from the confining potential}
Since the Hamiltonian density operator $\mathcal{H}_C(\bfvec{r})$ is not translationally invariant, the associated confinement energy can only be defined through a baryonic density defined in the Wigner sense, according to the justification given in the introduction: 
\begin{eqnarray}
 \varepsilon_C(r)&=&   \int\frac{V d^3\Delta}{(2\pi)^3}\, \langle \bfvec{\Delta}_N/2|\,\mathcal{H}_C(\bfvec{r}) \,|-\bfvec{\Delta}_N/2\rangle= W_C(r)\,\int\frac{V d^3\Delta}{(2\pi)^3}\, \langle \bfvec{\Delta}_N/2|\,q^\dagger(\bfvec{r})\,q(\bfvec{r})|-\bfvec{\Delta}_N/2\rangle\nonumber\\
 &=& 3\,W_C(r)\,\rho_B(r).\label{WIGN5}
\end{eqnarray}
And a similar definition for the ``bag pressure": 
\begin{eqnarray}
- p_C(r)&=& -3\,  \int\frac{V d^3\Delta}{(2\pi)^3}\, \langle \bfvec{\Delta}_N/2|\,\mathcal{P}_C(\bfvec{r})\,| -\bfvec{\Delta}_N/2\rangle=- r W'_C(r)\,\int\frac{V d^3\Delta}{(2\pi)^3}\, \langle \bfvec{\Delta}_N/2|\,q^\dagger(\bfvec{r})\,q(\bfvec{r})|-\bfvec{\Delta}_N/2\rangle\nonumber\\
 &=&- \, r\,W'_C(r)\,\rho_B(r).\label{WIGN6}
\end{eqnarray}
At this stage, it is worth discussing the purely numerical aspects of the computations. As already mentioned, the Wigner quark densities, appearing in Eqs.\eqref{WIGN1},\eqref{WIGN2},\eqref{WIGN3},\eqref{EQKW},\eqref{EQMW},\eqref{WIGN5},\eqref{WIGN6}, might be plagued by the oscillatory behaviour occuring  in the large-distance domain,  which is most likely an artifact of the delicate double numerical integration of the product of two Bessel functions. We first carefully tested the momentum integrations (and the double momentum integration for the various pionic energy densities) by checking their insensitivity to the number of Gauss points, the momentum cutoff, and the choice of the integration variable. We then checked that the sum rule, \(\int d^3r\,\rho_{B,S,K}(r)=1\), is satisfied to better than one part in a thousand for each of the three densities. We further tested the numerical integration of the densities using a Gaussian test density, $$\left(\psi^\dagger_E \psi_E\right)(x)=\left(\frac{1}{\pi \,b_E^2}\right)^{3/2}\,e^{-x^2/b_E^2},$$ for which the corresponding Wigner density can be obtained analytically: $$\rho_W(r)=\left(\frac{9}{ 4\pi\,b_E^2}\right)^{3/2}\,e^{-9\,r^2/4\,b_E^2}.$$ 
One can check that the numerically obtained Wigner density is indistinguishable from the exact one. To explicitly validate  the numerical integration, we also calculated the matrix element of a pure linear potential using this test density. The exact value is $\left\langle\sigma r\right\rangle=289.84\,MeV$, whereas the numerical results lie between $\left\langle\sigma r\right\rangle=289.70\,MeV$  and $\left\langle\sigma r\right\rangle=291.11 \,MeV$,   depending on the momentum cutoff and the integration variable. The latter case corresponds to the largest artificial oscillations. This serves as a numerical test for the worst-case scenario, where long-distance artificial oscillations are most likely to be amplified.\
\subsection{Nucleon mass}
If we first take as before the ``canonical" value of the string tension, namely $\sigma=0.18\,\rm{\rm{GeV}}^2$, and the pionic size parameter, $c_\pi=0.8$, we find that mechanical equilibrium occurs at a very small value of the parameter $b$, implying a core rms of about $r_B=0.37\,\rm{fm}$. This is due to the fact that the negative pion-induced pressure is too large; in addition, the mass of the nucleon is much too large. If the size parameter is increased to a value of $c_\pi=1.2\, (\rho_\pi=0.37\, \rm{fm})$, the quark core rms $r_B$ indeed increases, but the mass of the nucleon remains much too large. However, if one now looks at Eq.~(24) of our previous paper \cite{Universe2025}, one can easily convince oneself that the ansatz which disentangles the CSB potential (self‑energy kernel) and the confining potential (confining kernel) proposed in Eq.~(25) of \cite{Universe2025} overestimates the latter. One may then consider plausibly replacing the string tension by an effective string tension that rescales the size‑dependent part of the confining potential as well as the confining pressure (the ``bag pressure”). If one sets its value   to $\sigma_E=0.12\,\rm{GeV}^2$, one obtains in particular $b_{E/virial}=0.423\, \rm{fm}$ leading  to a quark core radius $r_B$  close to $0.45\,\rm{fm}$. Admittedly, the vacuum nucleon mass, $M_N=1.058\,\rm{GeV}$, remains too high, but one could reduce it by considering the effect of the perturbative gluon‑exchange, estimated to be of the order of $- 100\, \rm{MeV}$ \cite{Universe2025}. Note also that in the work of Simonov et al.~\cite{Simonov2002}, also based on the FCM but with a different approach, a value of $\sigma_E=0.09\,\rm{GeV}^2$ is phenomenologically preferred to the canonical one. The value of $g_A=1.21$ is also improved compared with the localized nucleonic state case.\\
To conclude this discussion, some comments are in order.  We have to acknowledge that we have limited ourselves to identifying, without implementing a systematic procedure, values of the parameters $\sigma_E$ and $c_\pi$ that are compatible with the constraints coming from the nucleon mass, the quark core size $r_B$ and $g_A$. A systematic study of these parameters, as well as of the various approximations involved, including the commonly used truncation to the $\Delta$ sector, would certainly be very interesting, but is beyond the scale of this present work and should deserve further study.
\begin{figure}
\includegraphics[width=0.45\textwidth,angle=0]{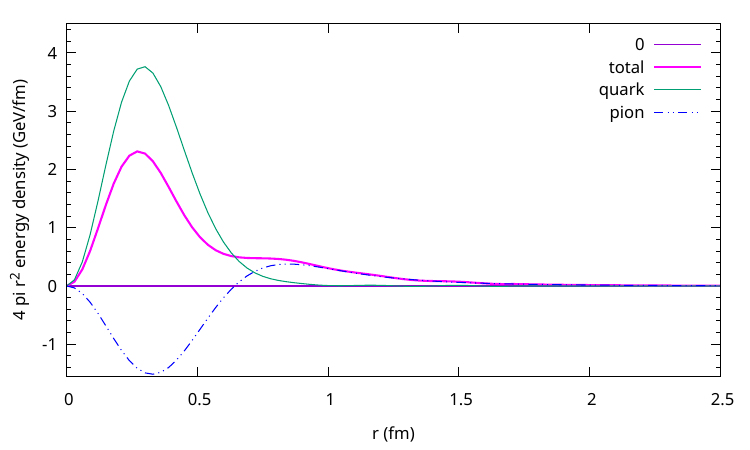}|
\includegraphics[width=0.45\textwidth,angle=0,]{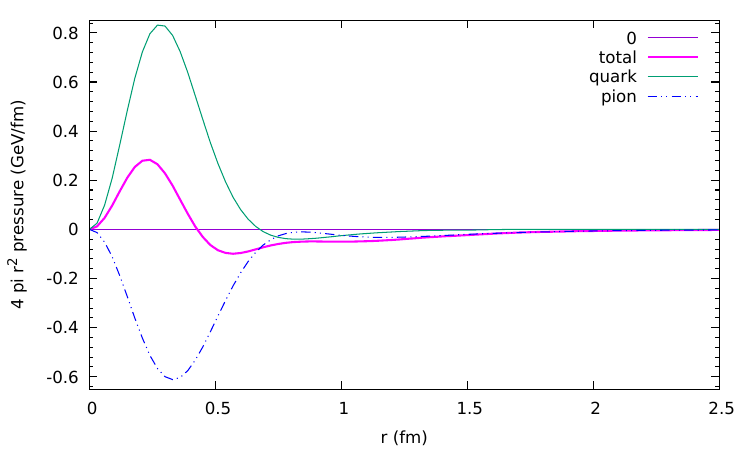}
\caption{Left panel: energy density distribution, $\varepsilon(r)=\varepsilon_{QK}(r)\,+\,\varepsilon_{QM}(r)\,+ \varepsilon_C(r)\,+\, \varepsilon_{\pi Q}(r) \,+\, 
\varepsilon_{\pi c}(r) \,+\,\varepsilon_{\pi K}(r) \,+\varepsilon_{\pi M}(r)$,  as a function of $r$ multiplied by $4\pi r^2$ for the physical nucleon state. Also shown are the separate quark and pionic distributions. Right panel: pressure distributions, $p(r)=\frac{1}{3}\varepsilon_{QK}(r)\,-\frac{1}{3} p_C(r)\,+\,\frac{1}{3} \varepsilon_{\pi Q}(r) \,+\, 
\varepsilon_{\pi c}(r) \,-\,\frac{1}{3}\varepsilon_{\pi K}(r) \,-\varepsilon_{\pi M}(r)$, as a function of $r$ multiplied by $4\pi r^2$. Also shown are the separate quark (i.e., non pionic) and pionic distributions. The effective  string tension is $\sigma_E=0.12\,GeV^2$  and the pion size parameter is $\rho_\pi =0.37\,fm$ ($c_\pi=1.2$).}
\label{CMENERGY}
\end{figure}
\subsection{Static EMT tensor: energy density and pressure}
\begin{figure}
\includegraphics[width=0.8\textwidth,angle=0]{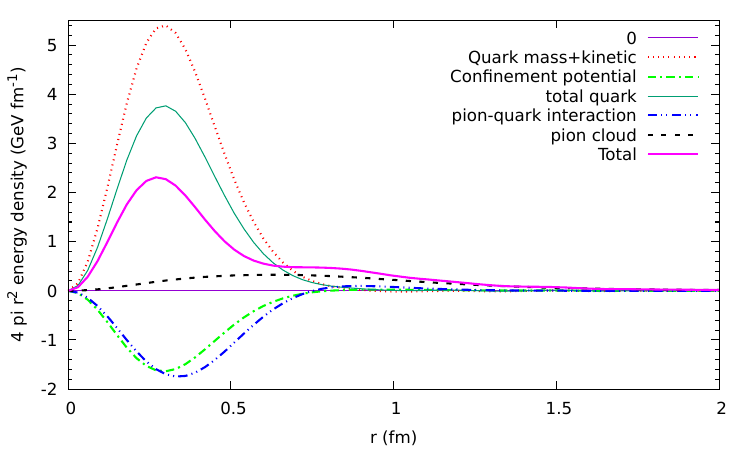}|
\caption{The various contributions to the energy density distribution for an isolated nucleon at rest. The effective  string tension is $\sigma_E=0.12\,GeV^2$  and the pion size parameter is $\rho_\pi =0.37\,fm$ ($c_\pi=1.2$).} 
\label{QUARKDENSITY}
\end{figure}

 The results of the calculation for the radial energy density and pressure  distribution are shown in Figs. \ref{CMENERGY} and \ref{QUARKDENSITY}.
It is also interesting to compare our results for the nucleon in vacuum with existing calculations. This allows us to verify that the pressure distribution is somehow different to that obtained with the chiral quark-soliton model presented in Fig. 4 of the Ref. \cite{Goeke}, where the magnitude of the pressure in the quark core is twice smaller than in our case.  As for the energy distribution presented in Fig. 1 of Ref. \cite{Goeke}, it is quite different in that it does not show, as in our case, two clearly separated regions associated with the quarks and the pion cloud. In fact, the result from that reference more closely resembles our results obtained with a lower constituent quark mass. Our result for the pressure distribution has a common feature with  the one presented in Fig. 7b of Ref. \cite{Lorce2019}, in the sense that the pressure in the central region is significantly larger than in the chiral quark model case. In this latter paper, the EMT components are obtained via a modeling of the gravitational form factors, directly from the generalized parton distributions (GPDs), which are accessible in several exclusive processes, such as deeply virtual Compton scattering \cite{DVCS} and meson production \cite{MESPROD}. An important point is that the energy distribution shown in Fig. 2b of this paper has a significantly larger extent than both the results from Ref. \cite{Goeke} and our own, despite the relatively long tail associated with the pion cloud in our approach.
\section{Summary and perspectives for the in-medium evolution of the nucleon properties }
In this paper, we have considered a class of nucleon models incorporating both a  confining potential and a standard pseudo-vector coupling of the constituent quarks to the pionic field. One important aspect  and original goal of this paper  is that the nucleonic wave function is constrained by the mechanical stability (von Laue) condition, which fixes the size of the quark core as resulting from a delicate balance between the positive Fermi pressure and the negative confining and pionic pressures.  All the formal developments of this paper concerning  the nucleon global properties (mass, size, axial coupling constant), as well as the internal energy and pressure distributions, will be generalized  in a companion paper \cite{companion} to the case  of the in-medium nucleon within the framework of the chiral confining model \cite{Universe,Universe2025}. In practice, this will be achieved using the same Lagrangian as Eq. \eqref{LAGEFF}
 but with parameters such as $M_q, F_\pi, M_\pi$ replaced by their in-medium analogs, $\mathcal{S}, F_\pi(\mathcal{S}) ,M_\pi(\mathcal{S})$. The background field $\mathcal{S}$, whose vacuum value coincides with the constituent quark mass $M_0$, is 
 the in-medium scalar field formally related to the ''nuclear physics'' sigma meson field, $\sigma_W$, of relativistic mean field theory through $s=(F_\pi/M_0)(\mathcal{S}-M_0)\equiv \sigma_W$, as explained in detail in Ref. \cite{Universe2025}.  
\section{Acknowledgment}
The authors acknowledge the support of the CNRS-IN2P3 MAC masterproject, the project RELANSE ANR-23-CE31-0027-01 of the French National Research Agency (ANR), the European Union’s Horizon 2020 research and innovation program under grant agreement STRONG–2020-No824093. The authors also acknowledge Jérome Margueron for valuable discussions and his continued interest in this work. One of the authors, GC, further acknowledges C. Lorcé and P. Schweitzer for highly informative email exchanges regarding the virial theorem and the static energy-momentum tensor. 
\appendix 
	
\section{Ansatz for the confining EMT operator and confining  stress tensor}
Let us introduce the following EMT operator:
\begin{equation}
\hat{T_C}^{i k}(x)=-\frac{1}{2} \,\mathcal{P}_C(x)\left(\delta^{ik}\,-\,\frac{r^i r^k}{r^2}\right). 
\end{equation}
$\mathcal{P}_C(x)$ is the "confining pressure  operator" introduced in Eq. \eqref{PCONFF},
$$\mathcal{P}_C(x)\equiv \hat{P}_C(x)=\,q^\dagger(\bfvec{r})\, \bfvec{r}\cdot \bfvec{\nabla}_{\bfvec{r}}W_C(\bfvec{r})\,q(\bfvec{r})=q^\dagger(\bfvec{r})\, r W'_C(\bfvec{r})\,q(\bfvec{r}),$$
where, again,  as in the rest of this discussion, the space-time point $x=(0,\bfvec{r}$) is implicitly taken at $t=0$. 
 For any (normalized) stationary state $|\Psi\rangle$, one has:
\begin{eqnarray}
 \langle\Psi| \partial_i \hat{T}_C^{i k}(\bfvec{r}) |\Psi\rangle &=& \-\langle\Psi|\partial_k W(\bfvec{r})\,q^\dagger(\bfvec{r})\, q(\bfvec{r})|\Psi\rangle\,-\,\frac{1}{2}\partial_i\left(q^\dagger(\bfvec{r})\, r W'_C(\bfvec{r})\,q(\bfvec{r})\right)\,\left(\delta^{ik}\,-\,\frac{r^i r^k}{r^2}\right).
\end{eqnarray}
The second term on the r.h.s. identically vanishes, since, for a spherical system, the gradient of the matrix element is proportional to $\hat{\mathbf r}_i$, whose contraction with the transverse tensor is zero.
It follows that:
\begin{eqnarray}
 \langle\Psi| \partial_i \hat{T}_C^{i k}(\bfvec{r}) |\Psi\rangle &=& \-\langle\Psi|\partial_k W(\bfvec{r})\,q^\dagger(\bfvec{r})\, q(\bfvec{r})|\Psi\rangle,\equiv -\langle\Psi| \partial_i \hat{T}^{i k}(\bfvec{r}) |\Psi\rangle .
\end{eqnarray}
Hence the divergence of $\hat{T}_C^{i k}(\bfvec{r})$ just compensates that of $\hat{T}^{i k}(\bfvec{r})$
according to Eq. (66). The corresponding confining stress tensor is simply
\begin{equation}
T_C^{i k}(\bfvec{r})=-\frac{1}{2} \,p_C(r)\,\left(\delta^{ik}\,-\,\frac{r^i r^k}{r^2}\right),
\end{equation}which satisfies $(1/3)T_C^{i i}(\bfvec{r})=-p_C(r)/3$. This is precisely the confining pressure distribution introduced in Eqs. (76-80) and whose explicit form is given by Eq. (123) for the localized state and Eq.(161) for the momentum projected state.\\
In a perspective of a future work, we note that by construction of the confining stress tensor $T_C^{ik}(\bfvec{r})$, the total static EMT should automatically satisfy $\partial_i\left(T^{ik}+T_C^{ik}\right)=0$, which translates into the local hydrostatic equilibrium equation,
$$\frac{dp_r(r)}{dr}=-\frac{2s(r)+p_c(r)}{r},$$
where $p_r(r)$ and $s(r)$ refer to the radial pressure and the shear force distribution of the original non-confining stress tensor, respectively. $p_C(r)=-2p_{tc}(r)$ is the correction arising from the confining stress tensor, which is of a purely tangential nature.


\begin{thebibliography}{999}
\bibitem{companion} G. Chanfray, H. Hansen, B. K. Pradhan, companion paper: "Mechanical properties of the nucleon in the chiral confining model. II - in-medium evolution of the nucleon properties"; arXiv : 2606.03633 [nucl-th] .
\bibitem{Universe} G. Chanfray, M. Ericson and M. Martini, Universe 9 (2023) 7, 316; arXiv: 2307.03484 [nucl-th].
\bibitem{Chanfray2024} G. Chanfray, Eur. Phys. J. A 60, 1 (2024); arXiv: 2310.19532 [nucl-th].
\bibitem{Universe2025} G. Chanfray, M. Ericson, H. Hansen, J. Margueron, and M. Martini, Symmetry 17, 313 (2025); arXiv:2501.10177 [nucl-th].
\bibitem{Tjon2000} Yu. A. Simonov, J. A. Tjon, Phys. Rev. D 62, 014501 (2000).
\bibitem{Simonov2002} Yu. A. Simonov, J. A. Tjon, J. Weda, Phys. Rev. D 65, 094013 (2002).
\bibitem{KNR2017} Yu. S. Kalashnikova, A. V. Nefediev, J. E. F. T. Ribeiro, Phys.Usp. 60 (2017) 7, 667-693; arXiv: 1707.04886 [hep-ph].
 \bibitem{Simonov2019} Yu. A. Simonov, Phys. Rev. D 99, 056012 (2019).
\bibitem{Gastao} G. Krein, Eur. Phys. J. A. 18, 511 (2003); M. E. Bracco, G. Krein, M. Nielsen, Phys. Lett. B 432, 258 (1998).
\bibitem{Wigner} M. Hillery, R.F. O’Connell, M.O. Scully, E.P.Wigner, Phys. Rep. 106, 121 (1984).
\bibitem{Lorce2019} C. Lorcé, H. Moutarde, A. P. Trawinski,  Eur. Phys. J. C,  79 (2019).
\bibitem{ChanfrayPRA} G. Chanfray and P. Schuck, Phys. Rev. A38, 4832 (1988). 
\bibitem{Schuck89} P. Schuck et al, Progr. Part. Nucl. Phys., 22, 181 (1989).
\bibitem{RGS} R. G. Sachs, Phys. Rev. 126, 2256 (1962).
\bibitem{Xiong}W. Xiong et al., Nature (London) 575, 147 (2019).
\bibitem{Polyakov} M.V. Polyakov, P. Schweitzer, Int.J.Mod.Phys. A33,  1830025 (2018). arXiv:1805.06596 [hep-ph]. 
\bibitem{DVCS} V.D. Burkert, L. Elouadrhiri, F.X. Girod, Nature 557  7705, 396 (2018).
\bibitem{MESPROD} L. Favart, M. Guidal, T. Horn, P. Kroll, Eur. Phys. J. A 52, 158
(2016); arXiv:1511.04535.
\bibitem{Brown} G. E. Brown, Nucl. Phys. A 358, 39c (1981).
\bibitem{Pirner} J. De Kam, H. J. Pirner, Nucl. Phys. A 389, 640 (1982).
\bibitem{KLE} S. P. Klevansky, Review of Modern Physics 64, 649 (1992).
\bibitem{vonLaue}M. von Laue, Ann. Phys. (Leipzig) 340, 524 (1911).
\bibitem{Polyakovor} M.V. Polyakov, Phys. Lett. B555, 57 (2003). 
\bibitem{Goeke} K. Goeke, J. Grabis, J. Ossmann, M.V. Polyakov, P. Schweitzer, A. Silva, and D. Urbano, Phys. Rev. D75, 094021 (2007). 
\bibitem{Lorce2021} C. Lorcé, A. Metz, B. Pasquini and S. Rodini, jheph11, 121 (2021).
\bibitem{Chanfray2007} G. Chanfray, M. Ericson, Phys. Rev. C 75, 015206 (2007).
\bibitem{CSN1990} G. Chanfray, P. Schuck and W. N\"{o}renberg,  Nucl. Phys. A519,  311c (1990).
\bibitem{Chanfray2011} G. Chanfray, M. Ericson, Phys. Rev. C 83, 015204 (2011).
\bibitem{Diakonov} D. Diakonov et al, Nucl. Phys. B480 (1996). 





\end{thebibliography}
\end{document}